\documentclass[12pt]{article}
\usepackage{amssymb}
\usepackage{amsmath}

\def\beq{\begin{equation}}\def\eeq{\end{equation}}
\def\bea{\begin{eqnarray}}\def\eea{\end{eqnarray}}

\begin{document}

\title{Pilot wave model without configuration or Fock spaces}
 
\author{Roman Sverdlov
\\Raman Research Institute,
\\C.V. Raman Avenue, Sadashivanagar, Bangalore --560080, India}
\date{December 6, 2010}
\maketitle
 
\begin{abstract}

\end{abstract}

The goal of this article is to come up with interpretation of quantum phenomena that is both local and deterministic. This is done by the means of envoking two different metrics, $g_o$ and $g_s$. These two metrics give very different "speeds of light": $c_o$ and $c_s$, respectively. The $g_o$ and $c_o$ are, respectively, "ordinary" metric and speed of light that we are used to. On the other hand, $c_s$ is superluminal. In this paper I propose a model in which newly introduced signals, which are subject to $g_s$, are responsible for key quantum phenomena. 

\subsection*{1. Introduction}

As was discussed in \cite{Africa1} and \cite{Africa2}, one of the biggest problems in interpretation of quantum mechanics is the need to envoke a configuration space. After all, if we only had one particle, we could have interpretted $\psi$ as merely a \emph{classical} field obeying Schrodinger's equation. Granted, the behavior of $\psi$ would have been "odd". In particular, for some strange reasons $\psi$ would happen to "collapse" into a $\delta$ function, the definition of "measurement" that causes a collapse would be unknown, and the location of collapse would be "probabilistic". However, as odd as the theory would be, we would still know what $\psi$ \emph{is}, even though we might not know why it behaves the way it does. 

Things become very different once we have more than one particle. In ordinary space, we can think of $\psi$ as the "amount of stuff" in a given location, whatever that "stuff" might refer to. In a configuration space this is no longer the case. Since we no longer have an intuitive picture of $\psi$, we are forced to call it "probability amplitude". This creates a paradox in itself: how can something related to "probability" be complex? Obviously, the way of removing that paradox is to go back to viewing $\psi$ as "physical" field. But this, again, brings back a question: if $\psi$ is physical, how can it be defined in a "configuration space"?

In this paper I propose to remove the concept of configuration space in favor of ordinary one. First of all, I will postulate a large set of point particles \emph{in the ordinary space} whose location is precisely defined in a \emph{classical} sense.  I will then postulate that some groups of them are "strongly correlated". Each "strongly correlated" group of particles \emph{in the ordinary space} corresponds to a \emph{single} particle in a "configuration space". Naturally, the numbers of particles in each of the "correlated groups" might differ. This naturally gives us a \emph{union} of different configuration spaces corresponding to all possible particle numbers. This, of course, is a \emph{Fock space}. 

I will then define $\psi$ as an internal oscillation of each \emph{individual} particle in the \emph{ordinary} space. Due to the strong correlation between the groups of particles, each two particles within the same group have nearly the same value of $\psi$ at any given time. 

This part, of course, violates relativity. However, I propose to have two \emph{differet} metrics: an ordinary one, $g_o$ (or, in component form, $g_{o; \mu \nu}$), and "superluminal" one, $g_s$ (or, in component form, $g_{s; \mu \nu}$). The speed of light ($c_s$) corresponding to $g_s$ is much larger than the speed of light ($c_0$) corresponding to $g_0$. In fact, a particle moving with the speed $c_s$ would circle the universe multiple times within unnoticeably short time period. 

Nevertheless, its $c_s$ is still finite. This allows us to impose "superluminal" correlations without sacrificing locality. Of course, the price for this is that the correlations are no longer exact. Thus, different particles within the same "group" might have slightly different values of $\psi$. Thus, the \emph{exact} value of $\psi$ corresponding to the "group" of particles is not defined; but that is okay since the latter is non local and, therefore, not physical. However, due to the fact that $c_s$ is very large, the fact that the values of $\psi$ within the particles of the same group are \emph{nearly} the same. This allows us to propose a model whose predictions are consistent with quantum field theory as we know it. 

I will then go one step further and incoprporate Pilot Wave model proposed by D\"urr et al (\cite{jumps}) into my theory. The key concept behind the Pilot Wave models is that, on top of wave function, there is a "beable" which corresponds to "classical" reality. Different Pilot Wave models propose different things to be "beables": it might be position of the particles, or their momenta, or a classical field. The common denominator of most of these models is that the probability amplitudes evolve according to ordinary quantum mechanical laws, while "beables" evolve \emph{strictly} in response to the influence of the former (according to some dynamics known as "guidence equation").

Accordoing to the proposal due to D\"urr et al (\cite{jumps}) the "probability amplitude" is viewed from the point of view of quantum field theory as opposed to quantum mechanics. Likewise, the "beable" is a quantum state in a Fock space. Thus, at any given point in time, we have non-zero probability amplitudes for \emph{all} states and \emph{despite that} only \emph{one} state is actually realized. From time to time the "jumps" occur between different states. If, at the time $t$, the "realized" state $\vert e \rangle$, then the probability that, at the time $t+ dt$, a realized state will be $\vert e' \rangle$ is $\sigma (\vert e \rangle, \vert e' \rangle) dt$ where 
\beq \sigma (\vert e \rangle, \vert e' \rangle) = \frac{(\langle \psi \vert e \rangle \langle e \vert H \vert e' \rangle  \langle e' \vert \psi \rangle )^{\dagger}}{\vert \langle \psi \vert e \rangle \vert^2} \eeq
where $x^{\dagger}$ is equal to $x$ if $x \geq 0$ it is $0$ otherwise. Now, of course, the above theory implicitly assumes Fock space in its traditional sense. In this paper, however, I "redo" this in terms of correlated systems of particles. Just like the probablitiy amplitude corresponds to the internal oscilation $\psi$ of each particle, I claim that the "choice" of a "beable" state corresponds to the internal oscillation $B$. Again, $B$ is "correlated" so that particles belonging to the same group have very similar values of $B$. At the same time, $B$ has \emph{extra} restriction: only one of the correlated groups has "large" value of $B$. All other groups have $B$ close to $0$. Thus, $B$ allows us to "highlite" the "realized" configuration. 

Finally, we will "add" gravity to this picture. In order to avoid the problems related to attempts to quantize gravity, we will instead propose that gravity is a \emph{classical} field. We are able to do this because the above picture "converts" non-gravitational quantum field theory into classical framework. Thus, adding gravity is just a matter of "coupling" two "classical" theories. Now, since only the end-product of the work is "classical", it is important that gravity is added only \emph{after} all of the non-gravitational fields are "converted" into classical ones based on my description.

The goal of the paper is to come up with quantitative dynamics that accomplishes the above goals. 

\subsection*{2. Desired final outcomes}

In light of the fact that the theory is quite complicated, it is important to summarize the end-results that I am trying to accomplish, and then in the Chapter 3, we will actually work through the details of the theory.

\subsection*{2.1 Definition of probability amplitude and its desired evolution (flat case)}

As was emphasized in introduction, one of the key components of the theory is that there are different groups of particles that "correlate". This, naturally, leads one to ask: what makes particles "join" into "groups"? According to the proposal, each particle has an integer-valued "charge". This "charge" is not to be confused with the one in electrodynamics: in fact the version of a "charge" proposed here has integer values $1$ through $N$, where $N$ is very large. Furthermore, \emph{all} particles posess this version of a charge, regardless of their electric properties or anything else.

Now, this "charge" allows the particles to emit and receive certain frequencies. Thus, the particles with the same "charge" are, effectively, "listening" to the same radio station, and this is what makes them correlated. Now, we will assume that the particles are distinguisheable and each particle has number. Thus, $q_k$ is a "charge" of particle number $k$. As was said in the introduction, $\psi$ corresponds to the \emph{internal} oscillation of each particle. The $\psi$-osciallation of particle number $k$ will be denoted by $\psi_k$. Therefore, the "correlation" of $\psi$-values can be defined as a statement
\beq q_a = q_b \Rightarrow \psi_a (t) \approx \psi_b (t) \eeq
where the approximation sign is due to finite value of "superluminal" velocity $c_s$. In the above equation, $t$ represents "quasi-global" time: in light of the fact that $c_s$ is superluminal but finite, we have a very good \emph{approximation} to a global theory. It is also important to notice that, as $t$ changes, the values of $\psi_a$ and $\psi_b$ "change together. That is, they are only approximately equal to each other \emph{if} taken at approximately the same time $t$. 

From the above picture we see that each configuration of particles corresponds to a given value of charge. Since the value of $q$ ranges from $1$ to $N$, we only have $N$ points in Fock space. In other words, Fock space is discrete. At the same time, however, our original spacetime is continuous. Thus, we propose a discrete Fock space in a background of a continuum ordinary space. Furthermore, when we add gravity into the picture, we will see that 

Now, apart from defining $\psi$, we would like to also define transition amplitudes, $H$, between different states. Essentially, we would like to look at the probability amplitudes of transitions between different "local" parts of each state and "multiply" them to get overall probability of transition. However, in light of the fact that the things we are "multiplying" are complex-valued, we have similar problem as we have with $\psi$. 

We would like to use the same trick as we did with $\psi$. Now, if $H$ is correlated between particles of the same "charge", then each value of $H$ will be associated with \emph{one} configuration. But, in light of the fact that $H$ represents a \emph{transition} between two statees, it should be associated with \emph{two} configurations. In order to do this, we will introduce a different kind of charge, $q'$, that also varies from $1$ to $N$. 

Thus, if a given particle, $k$, has charges $q_k = s_1$ and $q'_k = s_2$, then $H_k$ represents a probability amplitude of transition from the state $\vert s_1 \rangle$ to the state $\vert s_2 \rangle$. Thus, in order for $H$ to be "consistent", we need the following correlations: 
\beq (q_a = q_b) \wedge (q'_a = q'_b) \Rightarrow H_a \approx H_b \eeq
Now, as we will see from the following sections, the mechanism of "correlation" will involve emission and absorbtion of signals with various frequencies. Now, in light of the fact that oscillation involves changing the sign, it would be meaningless to say that a given signal has communicated $\psi=-1$ as opposed to $\psi =+1$. The only thing a signal can communicate is the \emph{amplitude} of oscillation, which is a positive quantity.

Therefore, in order to be able to communicate complex valued quantities, we need to introduce four separate \emph{real and positive} fields $\psi_1$, $\psi_{-1}$, $\psi_i$ and $\psi_{-i}$, and do the same for $H$. Then we can, respectively, define $\psi$ and $H$ as 
\beq \psi = \psi_1 - \psi_{-1} + i \psi_i - i \psi_{-i} \; ; \; H = H_1 - H_{-1} + iH_i - iH_{-i}\eeq
The above definition leads to a redundancy, since $\psi$ and $H$ remain unchanged under the transformation
\beq \psi_c \rightarrow \psi_c + k \; ; \; \psi_{-c} \rightarrow \psi_{-c} + k \; ; \; H_c \rightarrow H_c + l \; ; \; H_{-c} \rightarrow H_{-c} + l \eeq
However, since our goal in this paper is to be as "literal" as possible, I claim that the above eight fields are all physical, and the two "pictures" in the above transformation are not the same. But, it just \emph{happends} that our dynamics is dependent on "overall" $H$ and $\psi$. As a result, it \emph{happened} that the two pictures are dynamically indistinguisheable even though they are not the same. 

In order to have a more compact notation, let us now introduce two classes of sets of particles:
\beq S_k = \{ a \vert   q_a = k \} \; ; \; T_k = \{ a \vert   q'_a = k \} \eeq
If we can accept a level of approximation due to finiteness of "superluminal" $c_s$, we can assign a "common value" to the particles of the same set. Thus, $\langle \psi \vert s_k \rangle$ is approximated by $\psi (S_k)$ and $\langle s_k \vert H \vert s_l \rangle$ is approximated by $H (S_k \cap T_l)$. Thus, our desired dynamics is
\beq \frac{d (\psi (t, S_k))}{dt} \approx \sum_l i H (S_k, S_l) \psi (x_l) \eeq
Let us now make the above equation more precise. Suppose we have two sets of particles: $a_n$ and $b_{kl}$, where $k$, $l$, and $n$ go from $1$ to $N$. Suppose we choose them in such a way that $q_{a_n} = n$, $q_{b_{kl}} = k$ and $q'_{b_{kl}}=l$. Then, according to the above equation, the following approximation should hold:
\beq \frac{d \psi_{a_k}}{dt} \approx \sum_{k,l} i H_{b{kl}} \psi_{a_l} \eeq
The above approximation is independent of the actual choice of $a_n$ and $b_{kl}$ that meet the above criteria. This independence is due to the "correlations" that we have introduced. 

Let us now rewrite a dynamics in terms of real components. In light of the fact that both $\psi$ and $H$ has four components, our equations would be enormously complicated. So, in order to simplify them, we will notice a sum over parameter $c$ that only takes four values: $1$, $-1$, $i$ and $-i$. This can be described by a single restriction, $c^4=1$. Thus, in this language, 
\beq \psi = \sum_{c^4 =1} c \psi_c \; ; \; H = \sum_{c^4=1} cH_c \eeq
In this notation it can be verified that $iH \psi$ is given by 
\beq \sum_l i H(S_k, S_l) \psi (S_l) = \sum_{l=1}^N \sum_{c_1^4 = c_2^4 =1} c_1 H_{c_2} (S_k, S_l) \psi_{c_2/c_1} (S_l, t) \eeq
And, therefore, the evolution equation becomes
\beq \frac{d\psi_{c_1} (S_k, t)}{dt} \approx \sum_{l=1}^N \sum_{c_2^4 =1} H_{c_2} (S_k, S_l) \psi_{-ic_2/c_1} (S_l, t) \eeq
If we again take the selection of particles $a_n$ and $b_{kl}$ described earlier, then the above can be rewritten as
\beq \frac{d\psi_{c_1} (a_k)}{dt} \approx \sum_{l=1}^N \sum_{c_2^4 =1} H_{c_2} (b_{kl}) \psi_{-ic_2/c_1} (a_l, t) \eeq
One can notice in the above equation that $H$ is taken at a \emph{single} particle, \emph{not} two particles. This is due to the fact that each particle has two different charges. Thus, the combination of the \emph{two} charges of \emph{one} particle tells us the information about the \emph{two} states involved in transition: the "in" state, given by $q_k$ and the "out" state, given by $q'_k$. 

Once again, there are many different choices of particles that can be made with the same $q$ and $q'$. The above approximation \emph{simulteneously} holds for \emph{all} such choices. This is the key place where we take advantage of the "correlations". The actual dynamics that describes how these correlations arise will be given in detail in section 3.1. However, that section can be understood without reading the rest of chapter 2. 

\subsection*{2.2. Desired Pilot Wave model (flat case)}

As was stated in the Introduction, after "converting" the probability amplitudes into our framework (sec 2.1 and 3.1) we intend to do the same with Pilot Wave model (sec 2.2 and 3.2). We have also stated that we would like to choose the Pilot Wave model proposed by D\"urr (\cite{jumps}) as our basis. The reason for this is that, as we have seen, our Fock space only has finitely many states. Thus, we would like to have a discrete Pilot Wave model in order to describe it. This makes D\"urr et al the best possible choice.

However, for the sake of convenience of a reader, we would like to first summarise the proposal of D\"urr (sec 2.2.1) before showing how we "convert" this proposal into our framework (sec 2.2.2). Therefore, a reader who is already familiar with D\"urr proposal in "standard" case can move straight to sec 2.2.2. However, sec 2.2.2 is important and can \emph{not} be skipped if one wishes to understand the proposed "local" development of Pilot Wave model in sec 3.2, since sec 2.2.2 outlines the key ideas of my own contribution to the theory. 

Nevertheless, a reader that is only interested in probability amplitudes as opposed to Pilot Wave models, can skip chapter 2 altogether and only read sec 3.1. It must be said, however, that Pilot Wave model is important in order to understand gravity part of sec 3.3, since gravity will be "coupled" to a Pilot Wave model. Sec 3.1 is non-gravitational and can be understood without chapter 2.  

\subsection*{2.2.1 Review of D\"urr et al}

The key idea behind Pilot Wave model is that we have two \emph{separate} substances: a particle and a wave. The most traditional verision of that model is the one proposed by DeBroglie and Bohm. According to that model, a wave evolves according to Schrodinger's equation, while a particle evolves according to \emph{guidence equation}, 
\beq \frac{d \vec{x}}{dt} = \frac{1}{m} \; \vec{\nabla} \; Im \; ln \; \psi, \eeq
The position of a particle is known at all times, and it is called \emph{position beable}. It is easy to see that, if the above equation is obeyed, then the "classical" probability current $\rho \vec{v}$ coincides with the "probability current" of Schrodinger's equation. As a result, if the "classical" probability density $\rho$ coincides with quantum mechanical, $\vert \psi \vert^2$, it will continue to do so. In other words, $\rho = \vert \psi \vert^2$ is an "equilibrium point" of a theory. 

Of course there is a suddlety here: in light of time reversal, if a system can not "leave" equilibrium, it can not "enter" equilibrium either. However, we know from classical probability theory that time reversal no longer holds on "coarse grained" level. Thus, we can reasonably expect $\rho \approx \vert \psi \vert^2$ on some "larger" scale.

The precise definition of the statements that I have just made is still an unanswered question. But, for the purposes of this paper, it would suffice to say that this question is related to \emph{classical} probability theory. Thus, it is still true that the model adresses "quantum mechanical" part of the problem: the probability we have to deal with is real, not complex; the particle moves along the deterministic trajectory and the probability is simply a measure of our "ignorance" due to not knowing exact initial conditions. 

Now, in case of quantum field theory, things are not nearly as simple. Since the number of particles change, we can no longer think of a beable as a "position of particles". In light of this, several proposals have been made in recent years in comming up with new "beables" for quantum field theory, such as beables due to Struyve and Westman (\cite{minimalist}), Dirac see beables due to Colin (\cite{collin}), particle beables in extra dimension (\cite{novisibility}), and so forth. According to D\"urr et al, however, the "beable" is simply a highlighted quantum state. Thus, a "wave function" in quantum mechanics generalizes to a \emph{set} of probability amplitudes in quantum field theory, while a "particle" in quantum mechanics generalizes to a state that is "realised" at a given time. 

Now, the essential goal is the same as it was in non-relativistic case: introduce the dynamics for a beable that would reproduce desired probability $\vert \psi \vert^2$. Now, since we have discrete states, the only thing available to us is a "jump". Since the jump is discrete, it can no longer be described deterministically. Therefore, the dynamics involves probabilities of jumps (but these probabilities are "classical" and, therefore, are both positive and real).

Thus,  the desired statement of the dynamics is that if a state $\vert e \rangle$ is realized at the time $t$, then there is a probability $\sigma (\vert e \rangle, \vert e' \rangle)$ that the state $\vert e' \rangle$ will be realised at the time $t + dt$. Now, if we denote by $\rho$ the probability that a state $\vert e \rangle$ is realised at the time $t$, then $\rho$ obeys the following equation:
\beq \frac{d \rho (\vert e \rangle)}{dt} = \sum_{e'} (\sigma (\vert e \rangle, \vert e' \rangle) \rho (\vert e' \rangle) - \sigma (\vert e' \rangle, \vert e \rangle) \rho \vert e \rangle ) \eeq
Therefore, we have to compare this equation with the one for $d \vert \psi \vert^2 /dt$, and come up with the definition of $\sigma$ that would make the two match. It is easy to see that the latter can be expressed as
\beq \frac{d}{dt} (\psi^* \psi) = 2 \; Re \; \Big( \psi^* \frac{d \psi}{dt} \Big) \eeq
Now, if we express the above in the form of quantum states, and insert the "identity matrix", we get
\beq \frac{d \psi}{dt} = i <e \vert H \vert \Psi> = i \sum_{e'} <e \vert H \vert e'><e' \vert \Psi>. \eeq
Finally, we have to multiply the above by $\psi^*$ which is given by
\beq \psi^* = < \Psi \vert e> \eeq
This gives us 
\beq \frac{d}{dt} (\psi^* \psi) = 2 \; Im \; \sum_{e'} \; ( \langle \Psi \vert e \rangle \langle e \vert H \vert e' \rangle \langle e' \vert \Psi \rangle ) \eeq
Let us now compare this to our equation for $\rho$, which is 
\beq \frac{d \rho (\vert e \rangle)}{dt} = \sum_{e'} (\sigma (\vert e \rangle, \vert e' \rangle) \rho (\vert e' \rangle) - \sigma (\vert e' \rangle, \vert e \rangle) \rho \vert e \rangle \eeq
From our intuition, we know that if $\sigma (\vert e \rangle, \vert e' \rangle)$ is large, then $\vert e \rangle$ and $\vert e' \rangle$ should be very similar to each other. Thus, we can replace $\rho (\vert e' \rangle)$ with $\rho (\vert e \rangle)$ to get the following approximation:
\beq \frac{d \rho (\vert e \rangle)}{dt} \approx \rho (\vert e \rangle) \sum_{e'}  (\sigma (\vert e \rangle, \vert e' \rangle - \sigma (\vert e' \rangle, \vert e \rangle)) \eeq
By comparing this to the result that we got from evolution equation, we see that our desired result is 
\beq \sigma (\vert e \rangle, \vert e' \rangle) - \sigma (\vert e' \rangle, \vert e \rangle) \approx \frac{Im \; \langle \Psi \vert e \rangle \langle e \vert H \vert e' \rangle \langle e' \vert \Psi \rangle }{ \rho (\vert e \rangle)} \eeq
Now, we recall that, when we did the Bohm's case in the beginning of this subsection, we did \emph{not} prove that $\rho = \vert \psi \vert^2$. \emph{Instead} we have proven that \emph{if} $\rho = \vert \psi \vert^2$ happens to hold at a given point in time, \emph{then} it will continue to do so. Now, in the present case, we would like to be able to mimic the above argument. Therefore, we \emph{assume} that $\rho = \vert \psi \vert^2$ at a time $t$. In light of this, we can substitute 
\beq \rho (\vert e \rangle) = \vert \psi (\vert e \rangle) \vert^2 = \langle e \vert \Psi \rangle \langle \Psi \vert e \rangle \eeq
 in a denominator, which gives us
\beq \sigma (\vert e \rangle, \vert e' \rangle) - \sigma (\vert e' \rangle, \vert e \rangle) \approx \frac{Im \; \langle \Psi \vert e \rangle \langle e \vert H \vert e' \rangle \langle e' \vert \Psi \rangle }{ \langle e \vert \Psi \rangle \langle \Psi \vert e \rangle} \eeq
Now we have a freedom of defining $\sigma$ in different ways. We would like to utilize that freedom in order to make sure that $\sigma$ is positive. It is easy to see that the right hand side reverses signs under the rearrangement of $\vert e \rangle$ and $\vert e' \rangle$. Therefore, we can simulteneously keep $\sigma$ positive and satisfy the above equation if we identify $\sigma$ with $({\rm RHS})^{\dagger}$, where $x^{\dagger}$ is equal to $x$ for $x \geq 0$ and $0$ otherwise. Thus, we obtain
\beq \sigma (e', e) \approx 2 \; Im \; \frac{(\langle \Psi \vert e \rangle \langle e \vert H \vert e' \rangle \langle e' \vert \Psi \rangle )^{\dagger}}{\langle e' \vert \Psi \rangle \langle \Psi \vert e' \rangle} \eeq
where we have used the fact that denominator is positive in order to pull it out of $\dagger$. Now, since we have no other clues as to how turn the above approximation into exact one, we will simply postulate the above equation as an \emph{exact} dynamics. After all, $\rho = \vert \psi \vert^2$ is a \emph{consequence} of Pilot Wave model, and not hte other way around. Thus, we can always postulate that the latter is exact, at the expense of assuming that the former is approximate. This brings us to the final proposal made by D\"urr et al: \beq \sigma (e', e) = 2 \; Im \; \frac{(\langle \Psi \vert e \rangle \langle e \vert H \vert e' \rangle \langle e' \vert \Psi \rangle)^{\dagger}}{ \langle e' \vert \Psi \rangle \langle \Psi \vert e' \rangle} \eeq

\subsection*{2.2.2 Conversion of D\"urr et al into our framework}

We would like to modify the above model in two ways:

A. We would like to turn a stochastic model into a deterministic one

B. We would like to get rid of references to Fock space in that model and "simulate" it in our ordinary space by means of local, but superluminal, signals propagating with speed $c_s$. 

We will leave the actual proposal of such model to section 3.2, but for now we will just outline some of the key elements. In reality, these are simply something that we would \emph{like} to obtain as an end result of some very complicated model that will be discussed in 3.2. 

First of all, we have to define what it means for a "configuration" to be "selected" in terms of ordinary space. We have already seen in 2.1 how we can express $\psi$ in ordinary space, despite the fact that $\psi$ was originally meant to refer to "configurations". We will now use the same principle. In order to make the two problems similar, we will assume that, on top of $\psi$, there is \emph{another} function, $B$ on a Fock space. That function is equal to zero for all states except for the "highlighted" one. Then, during the "transition" from $\vert e \rangle$ to $\vert e' \rangle$, $B(\vert e \rangle)$ changes from large value to zero, while $B (\vert e' \rangle)$ does the opposite. 

Now, in light of the fact that one of our goals is determinism, we would like to be able to write a differential equation. This means that we can not have discrete jumps. Therefore, $B$ has some "in between" values. During the "transition" period it is quite possible that $B$ is non-zero for \emph{both} of the states. But, we would still like to make sure that \emph{most of the time} $B$ is cloase to zero for all states except one.

Furthermore, in much the same way as the "common" value of $\psi$ is only an approximation, what we have just said about $B$ is approximation as well. In other words \emph{even far away from the transition region}, $B$ is \emph{not} exactly zero for non-selected states; it is simply very close to zero. Thus, we can formulate what we have said so far as follows: 

At a \emph{usual} time $t$  (that is, far away from "transition") the following conditions are met: 

1) If $q_a = q_b$, then $B_a (t) \approx B_b (t)$

2) There is a \emph{unique} $q(t) \in \mathbb{N}$ such that $q_a$ is large.

The variable $q(t)$ undergoes discrete jumps. The above approximations do \emph{not} hold in the vicinity of these jumps. Therefore, the exact time of the jump can not be defined. There is a \emph{continuous} deterministic process that we \emph{interpret} as the above "jumps". The desired \emph{deterministic} theory implies that the probability of these jumps is consistent with D\"urr et al (\cite{jumps}) and is given by 
\beq \sigma (e', e) = 2 \; Im \; \frac{(\langle \Psi \vert e \rangle \langle e \vert H \vert e' \rangle \langle e' \vert \Psi \rangle)^{\dagger}}{ \langle e' \vert \Psi \rangle \langle \Psi \vert e' \rangle } \eeq
Another thing that is important to adress is creation and annihilation of particles. Suppose, for example, we have a process where an electron and positron annihilate into two photons. According to our proposal, there are \emph{four} particles at all times (for convenience, we will number electron and photon with $1$ and $2$ and we will number the two photons with $3$ and $4$). 

Now, it happens that $q_1 = q_2 = Q_1$ and $q_3 = q_4 = Q_2$. At time $t_1$, \emph{all} particles with charge $Q_1$ across the universe were highlighted. Thus, $B_k (t_1) = A_1$ whenever $q_k = Q_1$, and it is $0$ otherwise. On the other hand, at time $t_2$, the particles with charge $Q_1$ were no longer highlighted and, instead, particles with charge $Q_2$ became highlighted. Thus, $B_l (t_2) = A_2$ whenever $q_l = Q_2$ and it is $0$ otherwise. 

As a result of this "global" phenomenon, electron and positron were "highlighted" at $t_1$ while the two photons were highlighted at $t_2$. Since we only see the particles that are "highlighted", it \emph{appears} to us that electron and proton have annihilated. In reality, however, they continue to exist, just in "invisible" form. Likewise, it \emph{appears} to us that the two photons were "created". In reality, they always existed; they were simply "invisible". Thus, we can formulate it by defining a "matter density" as 
\beq \rho_k (x) =  \sum_k \int d \tau_k \; B_k (\tau_k) \delta^4 (x - x_k) \eeq

\subsection*{2.3. Addition of gravity}

\emph{WARNING: This section might be controversial. But the non-gravitational part of the paper (that is, sections 3.1 and 3.2) does NOT depend on it. Therefore, one of the options is to skip section 2.3 and go directly to 3.1 and 3.2. However, 3.3 picks up gravity again, and, therefore, might likewise be skipped}

In order to avoid the difficulties that come from quatizing gravity, I propose to view it as a strictly "classical" field. Now, while other things are also "classical" thanks to Pilot Wave model, gravity is \emph{even more} so. When it comes to other fields, we still had a probability ampitude $\langle \psi \vert s \rangle$. We simply re-interpretted it in classical terms. For gravity, however, we do not have $\psi$ to begin with. Instead, we use $\psi$ for non-gravitational fields and then \emph{after} we obtain the behavior of beables of these fields, we look at how these fields produce gravity.

More precisely, we would like to say that gravity is a solution of an exact Einstein's equation, where the source term is given through the coupling of particles with $B$-field.  In order to avoid black hole singularities, $\delta$-functions might be replaced with some finite approximation, $\tilde{\delta}$. Thus, we would \emph{like} to say 
\beq f (R_{\mu \nu} (x)) = \sum_k \int d \tau \; B_k (\tau_k) \tilde{\delta}^4 (x - x_k) \eeq
On the left hand sinde we wrote $f (R_{\mu \nu})$ instead of standard Einstein's equation for a reason. According to Bianchi identity, the violation of energy momentum conservation implies lack of exact solutions. However, the variation of $B$-field changes the energy-momentum associated with particles. 

This issue is not unique to my specific Pilot Wave model. Even in the classical Bohmian model for fixed number of particles, they might accelerate through "guidence equation". Then, in more advanced models, such as D\"urr's, their numbers are not fixed which makes the situation even worse: during the "jumps" explained in section 2.2, new partices can be created "out of nowhere".  

There are deeper reasons as to why we can't restore energy conservation. First of all, according to Pilot Wave model the influence is \emph{one way} from wave to particle. Energy conservation, on the other hand, demands two-way interaction. Furthermore, even if Pilot Wave model did conserve energy-momentum, we would have to count \emph{all} of the "sources" of energy-momentum, which would include wave. However, due to decoherence, the wave splits into "parallel universes". Thus, by counting the wave as one of the sources, we predict gravitational interaction between "parallel universes" which, obviuosly, is very bad.  

It might be hoped, however, that once we do make sure that the interaction between the particle and a wave is two-way, the particle will "kill off" the unwanted "parallel universes"; then we will be able to count the energy of the wave without any fear of unwanted interactions. In fact, one example of a theory where a particle "kills off" parallel universes was presented in \cite{novisibility}. However, that particular model is highly non-local and, therefore, does not restore energy conservation that we are looking for. 

As far as the current paper is concerned we will simply admit the violation of energy conservaton as given, and modify Einstein's equation in such a way that it has solutions for non-conserved sources. This is done by introducing an extra term that breakes gauges invariance. However, we can not simply impose gauge fixing term. If we did, then the Bianchi identity would imply that the \emph{divergence} of the difference of $T^{\mu \nu}$ and gauge fixing is zero: 
\beq \nabla^{\nu} (T_{\mu \nu} - (GF)_{\mu \nu}) = \nabla^{\nu} G_{\mu \nu} = 0 \eeq
This, however, does not imply that $GF=0$. If we assume that, due to quatnum field theory, $T_{\mu \nu}$ is conserved, the most information we will get is that gauge fixing term is conserved as well. But, obviously, conservation does not set it to zero. If we try to impose initial conditions on it, then, over a very large period of time (such as, for example, the age of the universe), the "slight" non-conservation of $T_{\mu \nu}$ would "build up", so that the "initial conditions" will no longer be relevent. And, unlike gauge theories, we do not have luxury of denying the small non-conservation; after all, that is what we are trying to adress to begin with.

In order to avoid the possibly-large unwanted impact from gauge-fixing term, we have to make sure that our correction term comes with a very small coefficient, $\epsilon$. Now, we know that if we set $\epsilon$ to $0$ we will have no solution. This means that if $\epsilon$ is small, we should expect some kind of singularity to arise. However, we also know that solution \emph{does} exist for the conserved source. Therefore, the source of singularity is precisely non-conservation of energy. This means that as long as energy non-conservation is smaller than $\epsilon$, there is no singularity.

Now, the good news is that the conservation of energy momentum can be predicted from quantum field theory, independently of gravity. Thus, if Pilot Wave model is a good approximation to quantum field theory, it would not produce significant violation to energy momentum anyway. The bad news, however, is that, as long as particles are viewed as beables, in light of their point size their impact on energy momentum is dramatic; which automatically means that their violation of energy momentum conservation is dramatic as well. This means that it is important to "average out" the energy-momentum of particles on a larger scale:
\beq \tilde{T}_{\mu \nu} (x) \approx \int_{n_{\delta} (x)} d^4 x' \; T_{\mu \nu} (x') \eeq
where $n_{\delta}$ stands for a $\delta$-neighborhood which is "small" on classical scale but "large" on quantum mechanical. In section 3.3 we will introduce a "local" mechanism by which the "quasi-local" averaging works. This averaging, by the way, is also important in order to avoid "black hole" singularities created by point particles; in fact, when we replaced $\delta$ with its finite version $\tilde{\delta}$, we have assumed that we have some means of producing $\tilde{\delta}$, through averaging. 

However, since the "true" Lorentzian neighborhood is a vicinity of light cone and, therefore, not compact, we have to break Lorentz covariance in order to define $n_{\delta}$. While we are not explicilty doing the above integration, the "local" mechanism that we introduce happens to violate relativity. The above issue might be a "deeper reason" for this. Nevertheless, for the purposes of this paper, we are okay with violation relativity, as long as we continue to preserve "locality". For that reason, I still included that mechanism.  

Once the "classical" theory of producing gravity is under control, then our problem reduces to coupling two classical theories. It is easy to make an intuitive argument that it might well have a solution. After all, at a hypersurface $t={\rm const}$, we \emph{already know} $g^{\mu \nu}$ (both spacelike and timelike components). Thus, we can substitute that "fixed" $g_{\mu \nu}$ into a Pilot Wave model to see how our system will evolve between time $t$ and $t+dt$. Then, we can use this information as a "source" of gravity which, through the modified Einstein's equation will give us $g_{\mu \nu}$ at $t+dt$. Then we will repeat that procedure between $t+dt$ and $t+2dt$, and so forth.

There are, however, two more problems. First of all, if we are "true" to general relativity, then the particles will have to move along geodesic. This conflicts with our usual notions. When we discretize ordinary space, the "lattice" we use is stationary. Since our model is a discretization of Fock space, and our "groups of particles" correspond to "lattice points" \emph{on a Fock space}, we have to make sure they are also stationary, by analogy. In other words, states, themselves, would move as opposed to different motionless states being highlighted at differnt times. 

This, however, does not dismiss the theory out of hand. After all, by uncertainty principle, if the "position" of particle is precisely defined, the momentum is infinitely undertain. Therefore, setting momentum equal to zero is just as "wrong" as setting a particle to move with any other velocity. Thus, since the former "works" the latter might "work" as well.  In fact, in some respect, it might even be "better" to define state in terms of concrete particles moving along geodesics (as "messy" as that picture might end up being) than to stick to a traditional view that in curved spacetime there is no such thing as "state".

 But, at the same time, it this would destroy \emph{flat} space quantum field theory, since now the particles are allowed to move in that setting, too. So the key question is whether or not the picture of "stationary lattice" (and in our case we have a "lattice" in Fock space) can be replaced with the one involving "moving lattice". Analysis of this issue is beyond the scope of this paper and is up to the verifiation in future research.

If it turns out that the particles should not move, we still have a way of salvaging our theory. Throughout the rest of the paper we will explicitly use two different metrics, $g_o$ and $g_s$, which, respectively, produce the speeds of signals $c_o$ and $c_s$. Now, $c_o$ is the "ordinary" speed of a signal that we are used to; on the other hand, $c_s$ is "superluminal". We account all of the seemingly-nonlocal processes in quantum mechanics to the signals propagating with speed $c_s$.

Now, when we write down geodesic equation for the motion of a particle, we have to use either $g_o$ or $g_s$ in a definition of Christoffel's symbols. Now, if we use $g_s$ and if we assume that the latter happened to be flat (after all, the only "gravity" that we have observed is the one involving $g_o$), then the particles will automatically go along straight line. Then, we can always postulate they are stationary without any conflict with geodesic equation. In this picture we can explain the \emph{appearance} of covariance with respect to $g_o$ in terms of a "coincidence" that Hamiltonian \emph{happened} to couple everything to $g_o$ in identical way, which made $g_o$ \emph{look} like geometry while, in reality, it is merely a \emph{field} on a background of $g_s$-based geometry. 

Another approach to get the same result, is to introduce a \emph{third} metric $g_l$ (here $l$ refers to lattice) for which speed of light, $c_l$ is much \emph{slower} than $c_o$. In this case the lattice points might "think" they are moving fast, when in fact, they move very slow ompared to any of the velocities we are concerned about, which is why they can be safely approximated as stationary. 

Another thing to be weary of when it comes to $g_o$ is the fact that particles might move "backward" in global coordinate $t$, such is the case with a black hole. This might lead to conflict with Pilot Wave models that evolve "forward" in time. It is possible, however, to attempt to avoid this issue by rewriting everything in terms of $t$. Thus, the geodesic equation is 
\beq \frac{d^2 x_a^k}{dt} = \Gamma^k_{00} + \Gamma^k_{0l} \frac{dx_a^l}{dt} + \Gamma^k_{ij} \frac{dx_a^i}{dt} \frac{dx_a^j}{dt} - \frac{dx_a^k}{dt} \Big(  \Gamma^k_{00} + \Gamma^k_{0l} \frac{dx_a^l}{dt} + \Gamma^k_{ij} \frac{dx_a^i}{dt} \frac{dx_a^j}{dt} \Big) \eeq
While this might look like cheating, this is, in fact, consistent with the spirit of interpretation of quantum mechanics where we assume absolute time and view the appearance of relativity is "coincidence" in Hamiltonian. Now, the above equation is postulated if we choose the option of sticking with $g_o$. On the other hand, if we stick with $g_s$ we would violate relativity in a lot more obvious way, described in previous paragraph.

Now, as far as this paper is concerned, I intend to leave a lot of freedom for future research to decide between $g_o$ and $g_s$ in different aspects of the theory. In these cases I will use other letters in place of $o$ or $s$. As far as this paper is concerned, all of these choices seem to "work" but they have very different ideological implications. So I would like to be pluralistic and allow all of them.

\subsection*{3. The proposed model}

\subsection*{3.1. Quantum field theory in flat space}

Note: in sections 3.1.1 -- 3.1.3 we will assume that $H$ is already known, and will be using it in order to obtain the dynamics for $\psi$. Then, in section 3.1.4 -- 3.1.6, we will describe how $H$ has been "produced". 

\subsection*{3.1.1 Qualitative outline of proposed model}

In this model we imagine that our universe is compact. Furthermore, the particles emit waves that propagate with "superluminal" speed $c_s$. The latter happens to be so large, that the waves can circle the universe multiple times with a negligeable time period. Particles can also "absorb" waves. But, the process of "absorbtion" is very unusual: when a particle "absorbs" a wave, it \emph{only} has an impact on a particle but \emph{not} on a wave! In other words, the wave continues to propagate with the same intensity after so-called "absorbtion". This allows this wave to be "absorbed" multiple times by several particles. 

Now, each particle has emission and absorbtion frequencies that, respectively, are equal to its "charges" $q'$ and $q$. Thus, \emph{if} a particle number $k$ was "triggered" to emit a wave, it would emit the one with frequency $q'$. On the other hand, if the wave happened to be propagating from some other source, and it \emph{happened} to reach a given particle, it will \emph{only} be absorbed \emph{if} its frequency \emph{happened} to be $q$.

The difference between emission and absorbtion frequencies accounts for the communication between different "configurations". Suppose we have two particles, $1$ and $2$. It is easy to see that if $q'_1 = q_2$, then the particle number $1$ will be able to emit a signal that will be received by a particle number $2$. But, \emph{since absorbtion has no effect on a wave} that signal will \emph{also} be absorbed by a particle $3$, satisfying $q_3 = q'_1$. In other words, it will be \emph{collectively} absorbed by \emph{all} of the particles belonging to the same "group" as particle number $2$. 

By the same token, \emph{all} of the signals that have ever been emitted are absorbed "together" by all the particles in the group $2$. Now, since the value of $\psi$ is produced through the absorbtion of signals, it is only logical that all of the particles in group $2$ have approximately the same value of $\psi$. The same, of course, is true for all other groups of particles. Now, in light of time delay due to finite value of $c_s$, the equality is not exact. 

In order for the approximation to be good, the time delay related to signal circling the universe should be negligeably small (that is, smaller than the time it takes for "local" processes to occur that we use as "watch"). Of course, if the universe was infinite, the time delay would have been infite too, regardless of how large $c_s$ might be. That is why we are assuming that the universe is finite, which allows us to claim that the signal moving with velocity $c_s$ can circle the universe within very short time. Furthermore, we assume that our universe is compact in order to avoid various problems that might arise "on the boundary". 

Now, in light of compactness of the universe, the signals "come back" multiple times. Since they are "massless", the signals emitted a million years ago have just as strong effect as the ones emitted just now! Of course, we do not want signals emitted millions of years ago to cause \emph{change} in $\psi$. This can be avoided by modeling the "absorbtion" of signals as a driven harmonic oscillator with slight damping. The "receptor" of a signal is the "oscillating" quantity, while the signal itself is "driving force". 

Thus, $\psi$ and \emph{not} its time derivative is a function of the signals that are "out there". Whenever new signal is "added", this amounts to increase in the value of $\psi_c$ for one of the $c \in \{1, -1, i, -i \}$ (and, therefore, changes the value of $\psi$) among the configuration of particles whose charges coincide with its frequency. If, on the other hand, a signal would "die out" that would amount to $\psi$ "reversing back" to its earlier stage. Since this does not happen, we do \emph{not} want them to die out, which is why that is \emph{good} that signals emitted millions of years ago have the same intensity as the ones emitted recently. 

Now, on a first glance one might worry that this leads to a prediction of steady \emph{increase} of $\psi$ over time, while in reality it both increases and decreases. This can be explained in terms of the "four fields" ($\psi_1$, $\psi_{-1}$, $\psi_i$ and $\psi_{-i}$). While it is true that the affected field \emph{in}creases, the total $\psi$ can go both ways. After all, if the \emph{in}creased field is $\psi_{-1}$ this would result in \emph{de}crease in $\psi$. 

Therefore, over time, the increases of $\psi_1$ and $\psi_{-1}$ "cancel out" which leads to $\psi$ averaging to the same thing. Nevertheless, it is true that the individual values of $\psi_1$ and $\psi_{-1}$ can serve as a "clock" of the age of the universe. Thus, when the universe was "young", we could have had $\psi = 5-5 = 0$, while for the "old" universe we have $\psi = 10000 - 10000 =0$. But, since our dynamics is based on total $\psi$, we do not have any physical access to that extra information.

\subsection*{3.1.2 Emission of signals}

The qualitative model presented in 3.1.1. has one gap. Namely, if the signals are being emitted continuously, there might be interference. We would like interference terms to cancel out by making sure that emission events are "random". But, since we would like to have a deterministic theory, we do not want "true" randmoness. Instead, we need to come up with a "random generator" that is ran by deterministic laws. 

The specific way of "triggering" the emission is not important, as long as it is random. In this paper, we will propose that the emission is triggered when a particle "collides" with so-called "a-particles". Now, since particles are point-like, there is no "true" collision. Instead, we have to claim that a-particle has a "density" field, $\rho_a$. This field is very short acting. Thus, it forms a "cloud" that "looks" like a particle but really has size. In other words, we can think of $\rho_a$ as a "density" of "non-point" particle that pointlike particle "formed" around itself.

Now, the "relativistic" version of that density is a current, $j_a^{\alpha}$. Just like $\rho_a$ is really a \emph{field} rather than density, in the same way $j_a^{\alpha}$ is a vector field. The source of that vector field is a \emph{true} $\delta$-function corresponding to point particle. We have to be careful though: since our universe is compact, $j_a$ might "accumulate" in the universe (similarly to $\psi$) which would ruin our purpose of having it large \emph{only} in a vicinity of a particle. We adress this issue by adding a "mass term" to the dynamics of $j_a$. 

Thus, if we have $N_a$ $a$-particles, and each of these particles has a trajectory $a_k (\tau_j)$, then the resulting equation of motion for $j_a^{\alpha}$ is
\beq \nabla_j^{\beta} \nabla_{j ; \beta} j_a^{\alpha} + m^2_{j_a} g_{j; \mu \nu} j_a^{\mu} j_a^{\nu} = \sum_{k=1}^{N_a} \int d \tau_k \; \delta^4 (x - a_k (\tau)) \eeq
where $j$ in $g_j$ refer to the choice of a metric. As we said earlier we have two different metrics: $g_o$ for "ordinary" signals and $g_s$ for superluminal ones. In our case, since the field is short lived it is not important which we choose to use. Thus, I am leaving it up to the philosophy of a reader and/or future research by simply using $g_j$ and leaving it an open option for future to use either $o$ or $s$ in place of $g$. 

As far as the $j$ in $\nabla_j$, this means that we define it in terms of $g_j$-based Christoffel's symbols. It is also understood that the indeces in $\nabla_s$ are "raised" and "lowered" by $g_s$, while the indeces in $\nabla_o$ continue to be raised and lowered through $g_o$:
\beq \nabla_s^{\alpha} f = g_s^{\alpha \beta} \nabla_{s; \beta} f \; ; \; \nabla_o^{\alpha} f = g_o^{\alpha \beta} \nabla_{o; \beta} f \eeq
and, finally, $d \tau_{k; j}$ is defined by using $g_j$:
\beq d \tau_{k; j} = \sqrt{g_{j ; \beta \gamma} \frac{dr^{\beta}}{d \tau} \frac{dr^{\gamma}}{d \tau} } \eeq
We are now ready to talk about the actual emission of a signal once our particle of interest is being "triggered" by $r$-particle. As we have said in sec 2.1, we break our complex field $\psi$ into four positive real ones: 
\beq \psi = \psi_1 - \psi_{-1} + i \psi_i - i \psi_{-i} = \sum_{c^4 =1} c \psi_c \eeq
where it is understood that $c^4 =1$ under the sum is equivalent to $c \in \{1, -1, i, -i \}$. Therefore, we need four different means of emitting each of these four fields. We have seen from sec 2.1 that the desired equation for $\psi$ is 
\beq \frac{d\psi_{c_1} (a_k)}{dt} \approx \sum_{l=1}^N \sum_{c_2^4 =1} H_{c_2} (b_{kl}) \psi_{-ic_2/c_1} (a_l, t) \eeq
Therefore, we would like this to factor into our "emittor". At the same time we would also like to include the interaction of a particle with $j_a$. Thus, we would like to couple the velocity of particle to $j_a$ and then couple the product to both $\psi$ and $H$. So our proposed coupling is 
\beq g_{e; \alpha \beta} j_a^{\alpha} \frac{dx^{\beta}}{d \tau_e} \sum_{d^4 =1} H_d (k, \tau_e) \psi_{-id/c} (k, \tau_e)\eeq
However, we would like to make sure that emission frequency coincides with the charge $q'_k$. Thus, we have to say that the collision with $a$-particle \emph{first} creates oscillatory process with frequency $q'$ and \emph{then} this process causes the emission of $\mu$-field. Since the value of $q'$ is specific to the particle, this has to be internal oscillations of the particle itself, and, therefore, should not be confused with oscillations of $\mu_{\psi_c}$.

 I define these internal oscillations of particle by $e_{\psi_c}$. These have to be triggered by interaction with $j_a$. This means that once $a$-particle has passed, they have to "die out" very soon. For this reason, I model $e_{\psi_c}$ as a damped harmonic oscillator, driven by $j_a$: 
\beq \frac{d^2 e_{\psi_c ;k}}{d \tau_e^2} = -q'_k e_{\psi_c ;k} - \lambda_e \frac{d e_{\psi_c ; k}}{d \tau_e} +g_{e; \alpha \beta} j_a^{\alpha} \frac{dx^{\beta}}{d \tau_e} \sum_{d^4 =1} H_d (k, \tau_e) \psi_{-id/c} (k, \tau_e)\eeq
We are now ready to define a dynamics of $\mu$-field with its "source" being $e$-field. While with $e$-field we were undecided which $g$ to use (hence we used $g_e$), for $\mu$ we know we should use superluminal one, $g_s$, since we would like $\mu$ to circle the universe multiple times in order to assure "consistency" of $\psi$. Thus for the sake of generality, we use two different $g$-s, and write the dynamics of $\mu$-field as follows:  
\beq \nabla_s^{\alpha} \nabla_{s; \alpha} \mu_{\psi_c} = \sum_{k=1}^N \int d \tau_{k; s} \; e_{\psi_c} (\tau_{k; s}) \; \delta^4 (\vec{x} - \vec{x}_k (\tau_s) ) \eeq
It should be noticed that we did \emph{not} put mass term for $\mu$. This time, for the reasons explained in sec 3.1.1, we \emph{want} $\mu$ to circle universe multiple times (which is what we were trying to \emph{avoid} in case of $j$ by putting the mass term). Therefore, \emph{unlike} the situation with $j$, we make $\mu$ massless.

\subsection*{3.1.3 Reception of signals}

In sec 3.1.2 we have discussed the mechanism of emission of signal $\mu_{\psi_c}$. We denoted it $\mu_{\psi_c}$ instead of $\psi_c$ for a reason. As was described in sec 3.1.1., when the signal is omitted it fills the whole space. But, at the same time, it only gets absorbed by \emph{some} particles, namely the ones whose charge coincides with its frequency. What defines $\psi_c (\vert s \rangle)$ is the internal oscillations of that specific group of particles. Thus, in order to get $\psi_c$ we have to define a mechanism of \emph{absorbing} $\mu_{\psi_c}$. 

Now, as we said in sec 3.1.1., the absorbtion of signal is a \emph{one way} process: it has influence on the particle but it does \emph{not} diminish the intensity of a signal. Therefore, we can view this process as a driven harmonic oscillator, with natural frequency $q$:
\beq \frac{dr_{\psi_c; k}(\tau_s)}{d \tau_r^2} = - q_k^2 r_{\psi_c; k} (\tau_r) - \lambda_r \frac{d r_{\psi_c; k} (\tau_r)}{d \tau_r}+ \mu_{\psi_c} ( x_k (\tau_{k;r})) \eeq
where we have used $\tau_r$ ("r" standing for "reception") instead of $\tau_s$ because, while it is crucial that the "transmission" of $\mu$ is superluminal, it is not important when it comes to "reception" (just like it was not important for "emission" either).

In the above expression we have introduced the damping $\lambda_r$ in order to avoid singularity. After all, in light of the fact that possible charges are discrete, the "received" frequency is \emph{exactly} equal to the charge (as opposed to "approximately"), which is why we would have had singularity if we didn't introduce damping. However, $\lambda$ is assumed to be very small, so its effect on anything else is negligeable. 

Now, as one can see, $r_{\psi_c ; k}$ oscillates between positive and negative values. That is why we did \emph{not} identify it with $\psi_c$, since the latter is strictly positive. Therefore, we need to find some mechanism of "extracting" the \emph{amplitude} from the oscillations. The latter, of course, is positive, as desired. We do that by postulating the following equation:
\beq \psi_{c; k} = \sqrt{q_k^2 r^2_{\psi_c;k} + \Big(\frac{dr_{\psi_c; k}}{d \tau_r}\Big)^2} \eeq
where, again, we used a special $\tau_r$ corresponding to $g_r$ for "reception" since we are not sure which $g$ to use. This last equation is, in fact, the final definition of $\psi_c$. It is easy to see that the extracted quantity is strictly positive. That is the ultimate reason why we decided to use four different fields to define complex valued function. 

\subsection*{3.1.4 Hamiltonian: qualitative introduction}

In the previous three subsections we have shown in detail the mehanism for evolution of $\psi$. However, we know from quantum field theory that the derivative of $\psi$ is a function of $H$. Therefore, if we did our job correctly, we must have used $H$ \emph{at some point}. In fact, we have! The factor of $H$ have shown up in the equation for the dynamics of $e_{\psi_c}$. 

However, it is not correct to assume that we know $H$. After all, as we have seen from section 2.1, in our reinterpretation $H$ is \emph{not} a Hamiltonian but rather an internal degree of freedom of a particle. In fact, when we used $H$ in the equation for $e_{\psi_c}$, we explicitly said that $H$ is a function of $k$, where $k$ is a number of a particle. This, again, reinforces the fact that $H$ is an internal degree of freedom of a particle.

The reason why $H$ is related to Hamiltonian is very similar to a reason why $\psi$ is related to probability amplitude. There is a global correlation between the values of $H$ among the particles with the same $q$ and $q'$. Thus, the "common" value of $H$ among these particles "encodes" a transition probability between the group of particles with "first" charge $q$ and the group of particles with "first" charge $q'$. 

The key point is that the common value of $H$ corresonds to the transition amplitude given by quantum mechanical Hamiltonian. This statement is non-trivial and can not be simply postulated. We have to come up with \emph{dynamics} of $H$ that would show that this is, in fact the case. After all, we had to introduce dynamics of $\psi$ in order to show that the "desired" evolution was "true" (even though in standard quantum field theory the evolution equation is simply postulated, just like $H$ is). 

The good news, however, is that the mechanism of "generation" of $H$ is very similar to the one of "generation" of $\psi$, which is evident from the similarities between the statements of both questions. Therefore, wherever possible, we will borrow the evolution equations for $\psi$ with appropriate modifications.

We know that in standard quantum field theory $H$ represents the transition between two \emph{global} states; that is, each "state" describes the entire universe. At the same time, the "resultant" probability amplitude $H$ is "produced" by "local" probability amplitudes, $h$. This feature is similar to classical probability theory where a probability of "complicated" event is a product of probabilities of "simpler" sub-events.  This time, however, the probabilities are "complex" which means that we need to find a mechanism for this feature to arise. 

 In order to understand this point, consider the following example. Suppose first state consists of electron and positron near the point $A$ and a $W$-boson near the point $B$.  On the other hand, the second state consists of two photons near the point $A$, and an electron-neutrino pair near the point $B$. Thus, in order to find the probability amplitude of transition between the two states, we have to first ask ourselves what is the probability of annihilation of electron-positron pair near point $A$ into two photons. We then have to ask ourselves what is the probability of a decay of $W$-boson near point $B$, and then multiply those.

Now, in light of the fact that our configuration space is descrete (which is due to finite number of possible "charges" particles might have), the electrons are "displaced" by finite distance from the two photons they are supposedly decaying into. The same is true for the $W$ boson, electron and neutrino. Thus, from the "continuum" quantum field theory, probability of either process should be strictly zero. We, therefore, assume that we are given "effective" quantum field theory according to which the probability is non-zero as long as distances are sufficiently small. Formulating such effective theory is beyond the scope of this paper. Here, we will just assume that we know what it is, and our only goal is to "convert" it from position-based Fock space formulation into the ordinary space framework we propose.

Now, the distances between photons and electron might differ from the ones between electron, neutrino and W-boson. As a result, the probabilities of these processes might be different as well. Suppose that electron, positron and two photons are so close to each other that the annihilation is "very likely". At the same time, the W-boson, neutrino and electron are displaced from each other sufficiently far that the decay of $W$-boson is "unlikely". Then, due to the latter, the product of the two probability amplitudes might be small. Thus, the \emph{global} transition will not occur. But since we are "not allowed" to do local transitions, the decay of the electron and positron will not happen either. 

Now, in terms of $h$ and $H$ the picture is the following: the $h$-variable does \emph{not} have global correlation. Thus, $h$ for electron, positron and photon is large, while $h$ for $W$-boson, electron and neutrino is small. But, at the same time, $H$ \emph{does} have global correlation. Therefore, $H$ for \emph{both} sets of particles has to be the same, and it \emph{happened} to be "small" in both cases. But, at the same time, the global correlation of $H$ is produced through local process; just like with $\psi$, the appearance of non-locality is due to the speed $c_s$ of the relevent signals.

This is similar to voting for president. Suppose a person $a$ is in favor of candidate $A$, while a person $b$ is in favor of candidate $B$. Suppose candidate $a$ got elected. In this case, a person $b$ \emph{continues} to favor candidate $B$ but, at the same time, he obeys candidate $A$. The candidate that each person favors is $h$, while the candidate each person obeys is $H$. Thus, the fact that people $a$ and $b$ continue to favor different candidates indicates that $h$ does not have global correlation. However, the fact that they obey the same person indicates that $H$ does, in fact have one.

Now, strictly speaking \emph{both} $h$ and $H$ are local. After all, person $b$ "learned" about the results of election through radio waves, which propagate "locally". Besides, person $b$ decided \emph{as an individual} to obey the elected candidate A, \emph{despite} the fact that he does not favor him. After all, if every single \emph{individual} choses to disobey the elected president, then there would be no means of enforcing his "election". And, the decision of \emph{individuals} to obey the elected president is \emph{local}, although this "local" quantity is subject to "global correlation".

Finally, the globally-correlated quantity $H$ (that is, a decision to obey a given person) is produced through non-correlated quantity $h$ (that is, voting). In the same way, in quantum field theory the "correlated" quantity $H$ that represents global probability of transition is produced through non-correlated quantities $h$ that represent "local" probabilities of transition. Therefore, in order to formulate the mechanism of generation of $H$ we have to do the following steps:

1) Formulate the mechanism of generation of $h$. This is, in itself, non-trivial since, as mentioned earlier, due to discreteness of Fock space the "nearby" particles are slightly displaced from each other. Since our intended theory is \emph{trully} local as opposed to quasi-local, we need some "local" mechanism of "gathering" quasi-local information into the same point in order to produce $h$. 

2) Find a superluminal, \emph{but local}, mechanism by which the correlated quantity $H$ is generated once $h$ is produced at all points. 

\subsection*{3.1.5 Mechanism of generation of h}

We would now like to come up with a concrete model of generating the $h$. As we have established in the previous section, while $H$ is global, $h$ is quasi-local. However, $h$ is not truly local either. This is due to the fact that the number of possible charges is very large but finite. Therfore, Fock space is discrete. Thus, the "transition" between two "local" elements of Fock space is quasi-local, which means that $h$ has to be quasi-local as well.

Since we would like to come up with \emph{truly} local theory, we need to come up with a "local" mechanism for quasi-local $h$ to arise. As we have indicated previously, both $H$ and $h$ are incoded in \emph{one} particle rather than two, because each \emph{one} particle has \emph{two} charges. Thus, both $H_k$ and $h_k$ has \emph{something} to do with transition from $\vert q_k \rangle$ to $\vert q'_k \rangle$. 

Thus, both $H_k$ and $h_k$ are the functions of some information regarding the particles $l$ satisfying $q_l = q'_k$. $H_k$ is a function of \emph{all} such particles, throughout the universe, while $h_k$ is only a function of "nearby" ones. Now, in order for the mechanism of production of $h_k$ to be \emph{truly} local, we need to introduce signals that would encorce it. 

We have already described such signals in previous section when we talked about $\mu_j$ and $\mu_{\psi_c}$. Therefore, we would like to come up with yet another $\mu$, namely $\mu_{\rho'}$. While the theory is similar, there are a couple of changes that need to be made. First of all, we have to take $H$-dependence out of the equation for the "emitter". Thus, our new equation is  
\beq \frac{d^2 e_{\rho' ;k}}{d \tau_e^2} = -q'^2_k e_{\psi';k} - \lambda_e \frac{d e_{\rho' ; k}}{d \tau_e} +g_{e; \alpha \beta} j_a^{\alpha} \frac{dx^{\beta}}{d \tau_e} \eeq
Secondly, in order to make sure that our theory is quasi-local as opposed to global, we have to make sure that $\mu_{\rho'}$ dies out over a very short distance (as opposed to $\mu_{\psi_c}$ that "circles the universe" multiple times). Thus, we have to introduce a mass term, where it is understood that the "mass" is very large. Thus, our equation for $\mu_{\rho'}$ is
\beq \nabla_s^{\alpha} \nabla_{s; \alpha} \mu_{\rho'} + m^2_{\mu_{\rho'}} \mu_{\rho'} = \sum_{k=1}^N \int d \tau_{k; s} \; e_{\psi_c} (\tau_{k; s}) \; \delta^4 (\vec{x} - \vec{x}_k (\tau_s) ) \eeq
The equation for "reception" $r_{\rho'}$ is the same as before. As before, there is an important trick: the "absorbtio mechanism" is based on $q_k$ while the "emission mechanism" is based on $q_k'$. This is what allows the desired "communication" between two different states which would produce $h$ (which is a function of \emph{both} of them). Anyway, after we change the letters to suit our current purposes, the equation for $r_{\rho'}$ becomes
\beq \frac{d^2 r_{\rho'; k}}{d \tau_r^2} = -q^2_k r_{\rho'; k} (\tau_r) - \lambda_r \frac{d r_{\rho'; k}}{d \tau_r} + \mu_{\rho'} (x_k (\tau_{k; r})) \eeq
And, finally, as before, in order to get $\rho'$ we extract amplitude from the oscillation of $r_{\rho'}$, which gives us
\beq \rho' = \sqrt{q^2 r^2_{\rho'} + \Big(\frac{dr_{\rho'}}{d \tau_r} \Big)^2} \eeq
Thus, while the value of $\rho'$ is \emph{inside} the "receiver" particle, it represents the total density of the "emittors". In other words, we can picture that each "point" particle produced a "finite size" particle around itself. Thus, a "point" receiver-particle "overlapses" with "finite sized" emittor particles. $\rho'$ tells us about the "density" of emmitor particles at the location of a "point" receiver-particle.

Now, the information about the behavior of $\rho$ \emph{should} give a given point particle the information about the distances to all the other particles. Thus, if a given particle "knows" our version of quasi-local field theory, it can "use" the \emph{local} behavior of $\rho$ in order to calculate quasi local information that it will then "substitute" into the equation for $h$ and thus "calculate" the value of $h$ it is supposed to have.

There is a bit of a difficulty, however. If a point particle is allowed to use only the values of $\rho$ itself as well as its first derivatives, then this information can be "mimiked" by many different configurations of particles. Thus, it is possible for 10 photons to "mimic" the two-photons in terms of the local densities at key points; as a result, electron and positron will annihilate into 10 photons since they will "think" they are annihilating into $2$.

In order to avoid this difficulty, we have to come up with some mechanism by which the particle "watches" the behavior of $\rho$ "over time" and analyses it to get the actual distances. After all, in light of the fact that the emission of $\rho$ is triggered by the interaction with $a$-particles, and the latter keeps changing, it is nearly impossible for two different configurations to "mimic" each other's behavior \emph{over time}, despite the fact that it \emph{is} possible for them to mimic each other \emph{at an instant}.

The problem, though, is that we do not want non-locality in time, just like we do not want one in space. Therefore, we need to come up with some \emph{local} mechanism by which a particle "remembers" what went on over time. Our goal is for that mechanism to be something of the form 
\beq \frac{d^2 h_c}{dt^2} = f \Big(\rho', \frac{d \rho'}{d \tau_h}, h_c, \frac{dh_c}{d \tau_h} \Big) \eeq 
The fact that the right hand side has $h$-dependence implies "non-linearlity" which might provide a key for the "memory" to occur. However, we have not found the specific way of writing down the function $f$ on right hand side. The definition of $f$ is one, and only, gap in the theory. But this is a very important gap: until we know what $f$ is, we can not hope to explicitly reproduce any of the results of quantum field theory. 

Nevertheless, the same can be said about "standard" quantum field theory as well: until we know $H$ we can't dream to reproduce the evolution of states. The difference between what we do here and the standard case is that in case of the latter $H$ is viewed as "fundamental" while in our case $f$ is fundamental and $H$ is derived from it. But, in either case, we have to "guess" what the "fundamental" equation is, and, until we do, we can not make any kind of predictions. 

However, it is possible to discuss the interpretation of quantum field theory without knowing what the Hamiltonian \emph{actually} is. In the same way, in this paper we can discuss the interpretational aspects of our theory without knowing what $f$ is. Therefore, the end product of our work is that we will be able able to say "the interpretation is clear, and now get to work". Actually finding $f$ is left for the future research. 

\subsection*{3.1.6 Is H a sum or a product?}

As we have explained in section 3.1.5, there are two quantities, $h$ and $H$. The former only has quasi-local correlation, while the latter has global one. As we promised before, now that we have obtained $h$ we will proceed in finding the value of $H$ through it. We have to slow down, however, and decide whether we want to take a "sum" or a "product". On the one hand, we know from quantum field theory that total Hamiltonian is an integral of Hamiltonian density; this suggests "sum". On the other hand, based on what we were saying so far, the Hamiltonian represents probability. This implies "product".

Let us, therefore, do a simple example. We have four identical particles at locations $x_1$, $x_2$, $x_1'$ and $x_2'$, all of spin $0$. We assume that $x_1' \approx x_1$ and $x_2' \approx x_2$, and we also assume that $q_1 = q_2$ and $q_1' = q_2'$. Thus, we are interested in transition probability from $\{ x_1, x_2 \}$ to $\{x_1', x_2' \}$. Now, if we denote the creation and annihilation operators on position space by $\phi^{\dagger}$ and $\phi$, respectively, then we can write down Hamiltonian as 
\beq H = h_1 + h_2 = k_1 (\phi^{\dagger} (x_1') \phi (x_1) + \phi^{\dagger} (x_1) \phi (x_1')) + k_2 (\phi^{\dagger} (x_2') \phi (x_2) + \phi^{\dagger} (x_2) \phi (x_2')) \eeq
where the difference between coefficients $k_1$ and $k_2$ is due to the difference between the distances within corresponding pairs of particles (the above expression can be easilly obtained by "discretization" of $\partial^{\mu} \phi^{\dagger} \partial_{\mu} \phi$). Now, it is easy to see that, up to the normalization factors, 
\beq \langle x_1 x_2 \vert H \vert x_1' x_2' \rangle = 0 \; ; \; \langle x_1 x_2 \vert H^2 \vert x_1' x_2' \rangle = k_1 k_2 \eeq
Thus, the "sum" comes from the first equation, while the "product" comes from the last one. However, product appears only through $H^2$ term. If we do take this term seriously, it would give us problems if we consider situation with more than one kind of particle, which would lead to interference between different "channels" of going from initial to final state.

That term, however, can be avoided if we realize that we are \emph{only} talking about probability amplitudes within a \emph{small} time interval $\delta t$. In fact, as we established earlier, due to the large but finite amount of "charges" our configuraiton space is discrete. This implies that the "transition time" should have small but finite value. That value should be approximated, at least to the order of magnitude, based on the total amount of "states" in fock space (or, equivalently, the number of available integer values of "charges"). Since the latter is very large, $\delta t$ is very small, which means that $H^2$ term gives neglegeable contribution.

In order for $\langle x_1 x_2 \vert H \vert x_1' x_2' \rangle = 0$ not to cause a problem, we would \emph{like} to have a process $\{x_1, x_2 \} \rightarrow \{x_1', x_2 \} \rightarrow \{x_1', x_2' \}$. This, however, is "not allowed": the first and the second state, respectively, imply that particles at locations $x_1$ and $x_1'$ have the same "charge" as the particle at $x_2$; this, trivially, imply that $x_1$ has the same charge as $x_1'$ which means that we can not have a state containing $x_1$ that does not contain $x_1'$ or visa versa.

In order to solve this problem, we introduce another particle at $x_2''$ which is "very close" to $x_2$. In other words, the distance between $x_2$ and $x_2''$ is \emph{much} smaller than the one between $x_2$ and $x_2'$. Thus, the transition probabilities between the two are of the order of $1$ \emph{despite} the smallness of $\delta t$, and we can think of them as \emph{essentially} the same point. Therefore, the process $\{x_1, x_2 \} \rightarrow \{x_1', x_2'' \} \rightarrow \{x_1', x_2' \}$ will have probability of the order of $\delta t$ of each step, which is what we want.

At first it might look troubling because the product of two transitions, each having probability of the order of $\delta t$, will have "total probability" of the order of $(\delta t)^2$, which is the exact thing we said earlier we were trying to avoid. This, however, stops being the issue once we consider finite time intervals, say, $t_1 \leq t \leq t_2$. Then the number of choices of \emph{one} transition is approximately $(t_2 - t_1) /  \delta t$, which the one of \emph{two} transitions is $(t_2 - t_1)^2/ (2 (\delta t)^2$. Thus, a probability that single trial will produce "single transition" is of the order of $\delta t$, and the number of choices of "single trial" is of the order of $1/\delta t$. At the same time, the probability that "double trial" will produce "double transition" is of the order of $(\delta t)^2$ and the number of different possible double trials is of the order of $1/(\delta t)^2$. Thus, in both cases the product is finite. 

Let us now go back to the original question: is $H$ a sum, or a product, or what? From what we have just seen, we \emph{do not care} about $H^2$ term. Therefore, we might as well make it our goal to \emph{forbit} a single-step process $\{1, 2 \} \rightarrow \{1', 2' \}$ althogether, and only allow $\{1, 2 \} \rightarrow \{1', 2'' \}$ where $2''$ is "very close" to $2$ (see earlier discussion). Now, the way we can "forbid" things is through "products": the \emph{consequence} of postulating a product is that 
\beq H (\{1, 2 \} \rightarrow \{1', 2'' \}) = h (1 \rightarrow 1') h (2 \rightarrow 2'') = \delta t \times 1 = \delta t \eeq
and
\beq H (\{1, 2 \} \rightarrow \{1', 2' \}) = h (1 \rightarrow 1') h (2 \rightarrow 2') = \delta t \times \delta t = (\delta t)^2 \eeq
and, since whatever is of the order of $(\delta t)^2$ is inconsequential, this will, effectively, "forbid" $\{1, 2 \} \rightarrow \{1', 2' \}$. It is important to notice that $(\delta t)^2$ term might come with "wrong" coefficients; in particular, this might be the case if we have many different "channels" to go from initial to final state, which the above equation does not take into account. That is why a very important part of the argument was our "indifference" regarding anything of the order of $(\delta t)^2$. This argument implies that our goal is to simply make sure that whatever is of the order of $(\delta t)^2$ \emph{continues} to be of that order, while the coefficients don't matter \emph{as long as} this is the case.

However, our goal is to come up with local mechanism for every non-local ingredient in our model. From the example with $\psi$, we have seen that the local mechanism is \emph{additive}. To our luck, we know that the logarithm of a product is a sum of logarithms. Thus, we would like to use \emph{logarithms} in our theory. But, unfortunately, we have already used $H$, instead of $ln \; H$ in the previous sections. So, for the purposes of notation, we will say that 
\beq \tilde{H} = ln \; H \eeq
Again, for the purposes of notation, we will further say that the "integral" of $h$ produces $\tilde{H}$, \emph{not} $H$. Thus, the "local probability" is $e^h$. Therefore,
\beq \tilde{H} = \sum h_k \; ; \; H = exp (\tilde{H}) \eeq
Now, if $h \approx ln \; \delta t$ at two different places, we would have $\tilde{H} \approx 2 ln \; \delta t$, which would lead to $H = exp \; \tilde{H}$ being of the order of $(\delta t)^2$. Therefore, we would like $h$ to be $ln \; \delta t$ at only \emph{one} place, while being finite everywhere else. The "one" place where $h$ is very large corresponds to $x_1 \rightarrow x_1'$, while all the other places correspond to $x_2 \rightarrow x_2''$. 

Now, from the above equation we see a peculiar pattern: whenever $h$ is "large", it is \emph{always} approximately equal to $ln \; \delta t$, \emph{without} any extra coefficient (unless, of course, there is a generic one)! All the "physics" is hidden in "much smaller" $ln \; k$ term which, upon exponentiation, becomes a \emph{coefficient} that determins transition probability. This is even more peculiar than the constant value of $ln \; \delta t$, since our intuition will probably tell us to disregard $ln \; k$ as "random fluctuations". However, as they say, "if there is a will there is a way". Since we have not defined $h$, we certainly have a freedom to do so in such a way that we avoid "random fluctuations" and make the "small term" important. 

Our freedom of defining $h$ is logically parallel to the freedom of defining Hamiltonian in quantum field theory. We do not need to know Hamiltonians of any fields in order to understand the concept of ordinary QFT. The specific Hamiltonians are merely being "pluged into" that overall "framework". In the same way, the point of this paper is to develop a "framework" for the new, superluminally-local, QFT, and, later, we will "plug" $h$ into this "framework". The key difference between what we are doing here and ordinary QFT is the presence of non-trivial transition from $h$ to $H$ in our case, which is what makes $h$, as opposed to $H$, an item to be "plugged into" our framework.

Now, in case of ordinary QFT they were using compactibility with an experiment as a basis of defining Hamiltonian. In our case, in light of the complexity of the theory, we were unable so far to make a proposal for $h$ based on an experimental data. We do, however, have some "clues" of how $h$ should "look like". The above expression with "large" logarithmic term and "small" $k$-term is one such clue. We do, therefore, know that \emph{when} $h$ will be produced, it will, in some approximation, take the above form.

This, of course, is a very vague statement. If $h$ has $ln$-term only at \emph{some point}, then continuity demands that there should really be a \emph{region} where $h$ looks like this. And, there should be a "transition region" where $h$ is, say, $1/2 \; ln \; \delta t$. That "smaller" value of $h$ should not "mess things up" since the situation is identical in "transition regions" around all other "large regions"; thus, upon "averaging out" all coefficients are affected in the same way. But, of course, this argument is very vague. The more precise picture would only emerge when we will give quantitative definition of $h$.

As we have said in the paragraph before last, this situation is not too different from ordinary QFT in its embriotic form. We do not have enough evidence to provide equation for $h$. We do, however, have some clues about its properties. And, the idea that the "intermediate" vales of $h$ in "transition regions" average out to the same thing is one such property. 

\subsection*{3.1.7 Mechanism of generation of H}

In the previous section we have seen that $H$ is, indeed, the "product", but $\tilde{H} = ln \; H$ is the sum. From what we have done with $\psi$ we see that sum, and \emph{not} a product is what we have a "mechanism" of generating. Therefore, in this section we will "copy" what we have done for $\psi$ in order to obtain $\tilde{H}$ and then, at the very end, we will set up $H = exp \; \tilde{H}$. 

As we have seen in the previous section, $\tilde{H}$ is obtained by summing $h$ (which the local probability is given by $e^h$).  We have already done one important part of the work: we "brought" quasi-local information into "one point" through "large-mass" signals. So, now $h$ is defined \emph{pointwise}, and our task reduces to "summing" the "pointwise" information. This means that what we do here is even closer copy to what was done for $\psi$. The "emitor" equation is the same, up to change of letters:
\beq \frac{d^2 e_{\tilde{H}_c ;k}}{d \tau_e^2} = - \omega^2 (q_k, q'_k) e_{\tilde{H}_c ;k} - \lambda_e \frac{d e_{\tilde{H}_c ; k}}{d \tau_e} +g_{\tilde{H}; \alpha \beta} j_a^{\alpha} h_{c;k} (\tau_{\tilde{H}}) \frac{dx^{\beta}}{d \tau_{\tilde{H}}} \eeq
However, the "propagation" equation has to be somewhat different. In case of $\psi$, due to the fact that $\mu_{\psi}$ is massless, it has infinite life, which causes $\psi$ to "accumulate" and, therefore, change in time. In case of $\tilde{H}$ we do not want it to change in time. This means that we would like to supply $\mu_{\tilde{H}}$ with some pass in order to prevent it from living "for too long". At the same time, however, we would like $H$ to be globally-correlated. 

Thus, $\mu_{\tilde{H}}$ has to be very small (much smaller than, say, $m_{\mu_j}$ or $m_{\mu_h}$). While $\mu_j$ and $\mu_h$ can only travel a short distance, we would like $\mu_{\tilde{H}}$ to be "light" enough to circle the universe several times. At the same time, we would also like it to be "heavy" enough in order for its lifetime to be very small (and, as said before, the two requirenments do not contradict each other since the "superluminal" speed $c_s$ is very large). On any event, for appropriately-chosen $m_{\mu_{\tilde{H}}}$, the dynamics for $\mu_{\tilde{H}}$ is the following:
\beq \nabla_s^{\alpha} \nabla_{s; \alpha} \mu_{\tilde{H}_c} + m_{\mu_{\tilde{H}}}^2 = \sum_{k=1}^N \int d \tau_{k; s} \; e_{\tilde{H}_c;k} (\tau_{k; s}) \; \delta^4 (\vec{x} - \vec{x}_k (\tau_s) ) \eeq
The equation for the "reception" of the signal is carbon copy of the ones we found earlier, 
\beq \frac{d^2 r_{\tilde{H}_c; k}}{d \tau_r^2} = -q^2_k r_{\tilde{H}_c; k} (\tau_r) - \lambda_r \frac{d r_{\tilde{H}_c; k}}{d \tau_r} + \mu_{\tilde{H}_c} (x_k (\tau_{k; r})) \eeq
which then, by our usual trick, gives us a final answer for $H_c$:
\beq \tilde{H}_c = \sqrt{q_k^2 r^2_{\tilde{H}_c} + \Big( \frac{dr_{\tilde{H}_c}}{d \tau_r} \Big)^2} \eeq
Now that we have produced $\tilde{H}$, we convert it into $H$ through
\beq H = exp \; \tilde{H} \eeq
As before, our ultimate answer should be in component form, for which we have a lot of freedom. As far as the expression of $\tilde{H}$ in terms of $h$, there is one natural choice: 
\beq \tilde{H}_a = \sum h_{k; a} \; , \; a \in \{1, -1, i, -i \} \eeq
As far as $H = exp \; \tilde{H}$,  there is no one best choice. In this paper we will randomly decide to set $H_1 = - H_{-1}$ and $H_i = - H_{-i}$, which gives
 \beq H_1 = -H_{-1} = \frac{1}{2} e^{\tilde{H}_1- \tilde{H}_{-1}} cos \; (\tilde{H}_i - \tilde{H}_{-i}) \; ; \;  H_i = -H_{-i} = \frac{1}{2} e^{\tilde{H}_1- \tilde{H}_{-1}} sin \; (\tilde{H}_i - \tilde{H}_{-i}) \eeq
\subsection*{3.2 Beables (flat case)}

\subsection*{3.2.1 Encoding the information about $\sigma$}

We would now like to come up with local mechanism of generating the Pilot Wave model described in section 2.2. That is, we would like to be able to "highlight" different states with $B$ (as described in sec 2.2) and then make "jumps" which "translate" into us "changing" the state we highlight. As was explained in sec 2.2, the probability that such a jump occurs within a time interval $dt$ is given by 
\beq \sigma (e, e') = \frac{(\langle \psi \vert e \rangle \langle e \vert H \vert e' \rangle  \langle e' \vert \psi \rangle )^{\dagger}}{\vert \langle \psi \vert e \rangle \vert^2} \eeq
where $x^{\dagger}$ is equal to $x$ when $x \geq 0$ and $0$ otherwise. As we said previously, our goal is to come up with a deterministic mechanism to "generate" the above probability of transition. However, logically, we can not start doing it until we know the \emph{local} definition of $\sigma$. We have two challenges here. First of all, $\sigma$ is a function of configurations, and, secondly, it is a function of \emph{two} different configurations. Both of these challenges are similar to the ones we had to face when we were dealing with Hamiltonian. So we will use some of the tricks we have learned there.

We already know how to deal with the first challenge. In fact, we have already defined $\langle \psi \vert e \rangle$ \emph{at} a point of our interest. Our problem is to define $\langle \psi \vert e' \rangle$ \emph{at the same} point. Since $\vert e' \rangle$ is different from $\vert e \rangle$, they can not possibly refer to the same collection of points. Since the collections of points are defined based on the \emph{same} charge, $q$, it is clear that they do not overlap. Thus, with original prescription, $\vert e' \rangle$ can not possibly be read at the same point. In other words, $\psi_k$ does \emph{not} give us the information we want.

 What comes to our rescue, however, is a charge $q'$. Thus, we introduce a \emph{different} internal degree of freedom, $\psi'_k$. The trick is that the "emitor" and "messenger" for $\psi'$ is the same as for $\psi$: it is $e_{\psi_c}$ and $\mu_{\psi_c}$ \emph{as opposed to} $e_{\psi'_c}$ or $\mu_{\psi'_c}$. At the same time, however, the mechanism of reception is based on $q'$, \emph{not} $q$: 
\beq \frac{dr_{\psi'_c; k}(\tau_s)}{d \tau_s^2} = - q'_k r_{\psi'_c; k} (\tau_s) - \lambda_r \frac{d r_{\psi'_c; k} (\tau_s)}{d \tau_s}+ \mu_{\psi'_c} ( x_k (\tau_s)) \eeq
Thus, due to the fact that $\mu{\psi_c}$ is the same, $\psi'$ "listens" to the same thing as one of the $\psi$-s. But, due to the fact that $r_{\psi'_c}$ is different the "one of the $\psi$-s" we have just mentioned is \emph{not} $\langle e \vert \psi \rangle$. Instead, it is $\langle e' \vert \psi \rangle$ which gives us the informatoin we are missing. Both $\psi$ and $\psi'$ are taken at the same point (namely, at a particle number $k$), as desired:
\beq \psi_k = \langle \psi \vert s_{q} \rangle \; ; \; \psi'_k = \langle \psi \vert s_{q'} \rangle \eeq
The other ingredient, namely the $H$-matrix, has been already "converted" into a local quantity in the previous. For our current purposes, we can just write it symbolially as
\beq \langle s_{q_1} \vert H \vert S_{q_2} \rangle = H_k \; , \;  q_k = q_1 \; , \; q'_k = q_2  \eeq
Thus, we obtain
\beq \sigma (S_{q_1}, S_{q_2 }) = \sigma_k (\tau) \eeq
where
\beq \sigma_k (\tau) = \frac{Im (\psi_k (\tau)  \psi'_k (\tau) H_k(\tau))^{\dagger}}{\vert \psi_k (\tau)\vert^2}. \eeq
This, in fact, "encodes" the probability of transition between two states within a \emph{single} particle number $k$. As with everything else, we have a large choice of $k$. The approximation will hold for \emph{all} particles with the same $q$ and $q'$ due to global correlations. 

\subsection*{3.2.2. Where does "probability" come from}

In the previous section we were able to "locally" define a "transition probability" $\sigma$. As we have extensively discussed in section 2.2, $\sigma$ is "classical" probability. That is, while "quantum mechanical" probability $H$ is complex, the "classical" $\sigma$ is both real and positive. However, we are not done yet. At this point, $\sigma$ is only a \emph{desired} probability; not the actual one. In order for this to be actual probability in a classical sense, we have to come up with a deterministic process and find out that if we don't know the initial conditions, the probability \emph{happens} to be $\sigma$. 

In this section we will attempt to come up with that process. We propose that the "transitions" between two states are triggered by the "collision" between one of the particles of interst, and an $a$-particle. Since both are point particles, there is no actual collision. Instead, there is an overlap with $j_a$. Therefore, we would like to impose some "threshold" $M$: if $g_{\gamma \delta} j_a^{\gamma} dx^{\delta}/d \tau$ is greater than $M$ then a signal will be sent that triggers a transition between the states $\vert q_k \rangle$ and $\vert q'_k \rangle$. 

We would like to make sure that the probability of exceeding $M$ is, in fact, proportional to $\sigma$. According to the above definition of collision, that probability is proportional to the "volume" of the "cloud" around $a$-particle that is defined by $j$-field exceeding the above threshold. We know that the $j$ is produced through Laplace's equation, 
\beq \nabla_j^{\beta} \nabla_{j ; \beta} j_a^{\alpha} + m^2_{j_a} g_{j; \mu \nu} j_a^{\mu} j_a^{\nu} = \sum_{k=1}^{N_a} \int d \tau_k \; \delta^4 (x - a_k (\tau)) \eeq
and the trajectory of $a$-particle is given by geodesic equation, 
\beq \frac{d^2 a^\chi}{d \tau_a^2} = \Gamma_{a; \eta \lambda}^{\chi} \frac{d a^{\eta}}{d \tau} \frac{d a^{\lambda}}{d \tau} \eeq
This might and might not lead to spherical distribution, depending on our choices of $g$. In particular, if we set $g_j = g_s$ and $g_a = g_o$ then this would, in fact, be spherical distribution. In other cases, we would have to count Lorentz contraction, which might complicate the calculation of probabilities due to the fact that we are "not allowed" to assume Boltzmann distribution, unless we know mechanism by which it is produced.

Regardless of the details of the above, however, it is clear from dimensional analysis that $\sigma$ is proportional to $1/r$. Of course, due to mass, there is also $e^{-mr}$ factor. But we can simply assume that the threshold is so high that in order for the information to be "relevent", $r$ has to be very small, which results in $e^{-mr} \approx 1$. Now, dropping the unknown coefficient, we see that the probability of the collision is proportional to $r^3$. 

Thus, in spherical case, we would like the "threshold radius" to be proportional to $\sigma^{1/3}$. Now since $j$ is inverseley proportional to $r$ and $M$ is linearly proportional to $j$, it means that $M$ should be proportional to $\sigma^{-1/3}$.  Thus, the "emission" criteria is
\beq g_{s ; \gamma \delta} j_a^{\gamma} \frac{dx^{\delta}}{d \tau_s} (\tau_{\rm emission})> \frac{1}{p_{\rm emit} \sigma_k (\tau_{\rm emit})} \eeq
Let us now provide a mechanism by which the "collision" triggers a "transition". As we said before, the "highlighted" particles are defined by the internal degree of freedom $B$. Therefore, if a "transition" is to happen, we are to make sure that $B$ of one group of particles changes from something large to zero, while $B$ of another group of particles does the opposite. This, of course, requires a dynamics on $B$.

We propose the following mechanism. There are two other internal degrees of freedom, $A_k$ and $C_k$. The degree of freedom $A_k$ is responsible for "starting" $B_k$, while the degree of freedom $C_k$ is responsible for "stoping" $B_k$. Thus, the dynamics on $B_k$ is defined as follows:
\beq \frac{\partial^2 B_K}{\partial \tau_k^2} = A_k (\tau_k) - B_k (\tau_k) C_k (\tau_k) \eeq
Now, if we want to make a "transition" from a group of particles with charge $q$ to the group of particles with charge $q'$, we have to simulteneously trigger $A$ among the particles with charge $q$ and, at the same time, trigger $B$ among the particles with charge $q'$. This can be done by a similar trick that we used in defining $\psi'$ in the previous section. Namely, we will use \emph{the same} emittor and messenger fields, $e_{AC}$ and $\mu_{AC}$ for two \emph{different} receivers $r_A$ and $r_C$. If these two receivers are being "tuned" to different frequencies, then, as we have previousy discuss in detail, this would do a trick. 

Now, the key element in the argument is that the \emph{common} "emittor" has to have frequency $q$. In order to impose the above threshold, that "emittor" has to be triggered by a step function. Thus, its dynamics is given by
\beq \frac{d^2 e_{k; AB}}{d \tau_k^2} = -q_k^2 e_{k; AB} - \lambda_{AB} \frac{de_{k;AB}}{d \tau_k} + {\rm step} \Big(g_{s ; \gamma \delta} j_a^{\gamma} \frac{dx^{\delta}}{d \tau_s} - \frac{1}{p_{\rm emit} \sigma_k (\tau)} \Big) \eeq
The dynamics of messenger is defined based on emittor by our usual formula. This time we would like the "messenger" to be massive since we don't want a transition to "continue" to occur beyond the very short period of time after it has been "triggered". At the same time, we \emph{do} want the signal to circle the universe since the transition is global one. As we said before, due to very large value of $c_s$ it \emph{is} possible for a signal to be short lived \emph{and} circle the universe. But, what it menas is that its mass should be a lot smaller than short-range signals ($\mu_j$ and $\mu_h$) but at the same time much larger than long lived ones ($\mu_{\psi}$). Anyway, for appropriately chosen mass, the equation is 
\beq g_s^{\alpha \beta}  \nabla_{s ; \alpha} \nabla_{s; \beta} \mu_{AC} + m_{\mu_{AC}}^2 \mu_{AC} = \sum_{k=1}^N \int d \tau_{s;k} \; e_{k;AB} (\tau_k)\delta^4 (x-x_k (\tau_{s;k})) \eeq
Now, since we hae used $q_k$ instead of $q'_k$ in the equation for $j$, we are assuming that we would like to make a transition \emph{from} a configuration defined by $q_k$ \emph{to} a configuration defined by $q'_k$. Thus, if $q_l =q'_k$, then we would like to activate $B_l$ and de-activate $B_k$. This can be done by activating $A_l$ and $C_k$. Since $C_k$ happened to have "the same" index $k$, this means that the equation for the receptoin of $C$ is based on "the same" $q$:
\beq \frac{d^2 r_{C;k}}{d \tau_s^2} = -q'^2_k r_{C; k} (\tau_s) - \lambda_C \frac{d r_{C; k}}{d \tau_s} + \mu_{C;k} (x_k (\tau_{k; s})) \eeq
Since the key of our argument is the difference between the mechanism of reception of $C$ and $A$, we immediately know that the reception of $A$ has to be based on $q'$:
\beq \frac{d^2 r_{A;k}}{d \tau_s^2} = -q^2_k r_{A; k} (\tau_s) - \lambda_A \frac{d r_{A; k}}{d \tau_s} + \mu_{A;k} (x_k (\tau_{k; s})) \eeq
Finally, the actual values of $A$ and $C$ are given by usual means,
\beq A = \sqrt{q^2 A_r^2 + \Big(\frac{dA_r}{d \tau_s} \Big)^2} \; ; \; C = \sqrt{q'^2 A_r^2 + \Big(\frac{dA_r}{d \tau_s} \Big)^2} \eeq 
And, as was previously stated, the above are substitutted into a dynamics for $B$,
\beq \frac{\partial^2 B_K}{\partial \tau_k^2} = A_k (\tau_k) - B_k (\tau_k) C_k (\tau_k) \eeq
Now, as we explained in section 2.2, we identify "reality" with a probability density 
\beq \tilde{\rho}_k (\vec{x}, t) = B (\vec{x}, t) \sum_k \delta^3 (\vec{x} - \vec{x}_k) \eeq
Thus, the variation of $B$ might lead to the perceived creation and annihilation of particles. The "creation" and "annihilation" of similar particles nearby might lead to perceived motion, and so forth. In other words, the above gives us a complete description of classical reality.

\subsection*{3.3 Gravity}

\subsection*{3.3.1 Introduction}

We are now done talking about non-gravitational part of the theory and we are now ready to couple it to gravity. In section 2.3 we have introduced the basic concept of "coupling" classical general relativity with Pilot Wave model. We have stated, however, that non-conservation of energy momentum is a major issue. Therefore, we will focus this chapter on adressing that single issue. 

As we said in section 2.3, we would like to find a modification to gravity in such a way that it would be able to accomodate the violation of energy momentum conservation. However, we have pointed out that the fact that the Einstein's equation in its original form has \emph{no} solution for non-conserved case, the modified Einstein's equation has nearly-singular solution provided that its modification is "small" (we have also explained why "large" modifications through gauge terms are not acceptabe as they lead to "large", albeit devergence-less, violations of Einstein's equation).

We have pointed out, however, that in case of ordinary Einstein's equation the \emph{conserved} sources \emph{do} have a solution. Thus, the only source of "singularity" is precisely violation of conservation of energy momentum. Thus, \emph{if} we make sure that the violation of energy momentum conservation is much smaller than the "small" correction to Einstein's equation, this would not lead to singularity. 

However, we have observed that, in light of zero size of particles, any effect they have on gravity, including the violation of conservation of energy momentum, is arbitrarily large. Thus, we need "average out" energy momentum over sufficiently large volume in order to have any hope that the violation of its conservation is, in fact, "small". Appart from that, of course, they create unwanted black holes \emph{even if} the energy momentum tensor was conserved. This is another reason for having to average it out. 

In light of this, in section 2.3 we have proposed to adress the issue in two major steps:  

a) Average out energy momentum tensor over a sufficiently large neighborhood, in order to make sure that its non-conservation is very small

b) Propose modification of gravity that would have a solution for that very small violation of conservation

We will, therefore, do step a in sec 3.3.2, and step b in sec 3.3.3

\subsection*{3.3.2 Averaging out energy momentum tensor}

We will now attempt to propose a "local" way of "averaging" energy-momentum tensor. That is, we would like to take "large-fluctuating" $T_{\alpha \beta}$ and then produce "reasonable" $\tilde{T}_{\alpha \beta}$. In the previous sections, we have introduced various "messenger fields" $\mu$ for various other reasons. Since the "messengers" $\mu_{\rho}$ happened to obey regular wave equation, we can easily define energy-momentum tensor based on them. THe latter, of course, will be singular at the emission point, but we will soon be describing our ways of "converting" it into non-singular tensor.

However, in light of the fact that gravity is supposed to be produced by "beables", only the "highlighted" particles can emit our \emph{current} version of $\rho$ (which should be contrasted with all previous $\rho$-s which are produced both by highlighted and non-highlighted particles). Thus, this time we have to multiply the "emittor" by $B$. 

Furthermore, when we change from state to state, each new set of particles gravitates "in the same way". In fact, we are able to "observe" the gravity that was "emitted" in the past, even though back then a different state was highlighted. This means that the emission frequency has to be independent of $q$ and $q'$, so we have to remove this part from our equation. Thus, with the above two modifications, the "emittor" equation is 
\beq \frac{d^2 e_{T ;k}}{d \tau_e^2} =  - \lambda_e \frac{d e_{T ; k}}{d \tau_e} +g_{e; \alpha \beta} B_k (\tau_e) j_a^{\alpha} \frac{dx^{\beta}}{d \tau_e} \eeq
where we have retained the "damping" term in order for the gravity-production event to be limitted in time.  Since $e$ "takes care" of the above modifications, the equation of propagation of $\mu$ is similar to the one used previously, namely
\beq \nabla_s^{\alpha} \nabla_{s; \alpha} \mu_T + m^2_{\mu_T} \mu_T = \sum_{k=1}^N \int d \tau_{k; s} \; e_{T;k} (\tau_{k; s}) \; \delta^4 (x - x_k (\tau_s) ) \eeq
Now, since we would like to "average" our energy-momentum tensor over "large enough" region, $m_{\mu_B}$ has to be "small enough" for this to happen. In particular, it must be smaller than $m_{\mu_{\rho}}$ since the "size" of a particle (which we identify as a size of a "cloud" formed by $\rho$) is quantum-mechanical, while the size of "smearing" of energy momentum tensor is classical. In fact, I would like the gravity of particles making up a large object to "smear" so much that we see a "continuous" source of gravity which, naturally, is conserved.

 At the same time, however, $m_{\mu_B}$ has to be \emph{large} enough in order to make sure that the region $\mu_B$ fills up is \emph{small} on a \emph{classical} scale. Thus, the mass of $\mu_B$ is unusual compared to all the masses we took earlier. In the past, any messenger was either "heavy" enough to be constrained in a very small region or "light" enough to circle the universe. This time, however, we would like an in-between version, where it is light enough to fill a macroscopical region \emph{as long as} that region is "small" on our scale. 

Now, we define the energy momentum associated with $\mu_B$ by mimicking the way it is usually done for scalar fields: 
\beq T_{\alpha \beta} = \partial_{\alpha} \mu_{\rho} \partial_{\beta} \mu_{\rho} - \frac{1}{2} g_{o; \alpha \beta} g_o^{\gamma \delta} \partial_{\gamma} \mu_{\rho} \partial_{\delta} \mu_{\rho} \eeq
We have a problem now, since in the vicinity of a particle $T_{\alpha \beta}$ blows up. One thing we can do is to put an "upper bound" on $T_{\alpha \beta}$ by replacing it with $T'_{\alpha \beta}$ defined as follows:
\beq T'_{\alpha \beta} = M_0 T_{\alpha \beta} {\rm exp} \; \Big(- \sum_{\mu \nu} T_{\mu \nu}^2 \Big) \eeq
where we have violated relativity by assigning the same sign to \emph{all} terms in the exponent. This violation of relativity logically relates to the fact that Lorentzian neighborhood has infinite volume and, therefore, relativity has to be violated in order to define a "small" neighborhood. The reason for infinite volume of Lorentzian neighborhood has to do with difference in sign of a metric, and this is precisely what we got rid of in the above equation.  

However, we now have a different problem. What if we \emph{want} $T_{\alpha \beta}$ to be larger than the "upper bound" $M$ in the above equation? One way out of this situation is to say that the upper bound is only imposed on how much \emph{one} particle can contribute, since that is the ultimate source of singularity. But, as long as we are summing contributions from \emph{different} particles, we can have the value of $T_{ \alpha \beta}$ as large as we want. 

This, however, raises a question: the physics is not "smart", so how can it "know" where $T_{\alpha \beta}$ comes from? What comes to rescue is the emission process of the particles that is being "triggered" by various "random parameters" (which, in our case, is collision with $a$-particle). If we consider the case of $\mu_{\rho}$ field, due to its very large mass, it is highly unlikely that two different particles will emit $\mu_{\rho}$ so close in time that it would "add". Thus, we can safely interpret $\mu_{\rho}$ as a field produced by \emph{one} particle, and, therefore, impose our upper bound on it.

Now, we still want the version of $\tilde{T}_{\alpha \beta}$ that is produced by several particles. We can obtain it by comming up with a dynamics that would \emph{slowly} contribute to $\tilde{T}_{\alpha \beta}$ from the resources of $T'_{\alpha \beta}$. Thus, the former corresponds to the "integral" over some time interval. At the same time, that time interval can not be too large, in order to allow $\tilde{T}_{\alpha \beta}$ to change over \emph{sufficiently large} periods of time. We, therefore, propose the following dynamics:
\beq \frac{\partial \tilde{T}_{\alpha \beta}}{\partial t} = k_1 T_{\alpha \beta} - k_2 \tilde{T}_{\alpha \beta} \eeq
This dynamics violates relativity, as it singles out $t$ as "preferred" time direction. But, as we said before, the violation of relativity is linked to the non-local nature of \emph{true} Lorentzian neighborhood, which is defined by a light cone. If the above process went on \emph{all} timelike directions, it would, in particular, go along near-lightlike one. So by going with near-lightlike velocity in $+t/+x$ direction, and then "comming back" at $+t/-x$ direction, we would have arbitrary large effects arbitrary far in time. This means that if we did want to allow multiple time directions, we would be forced to impose a "cut off". That cut off would violate relativity. So, since we have to violate relativity anyhow, we might as well just consider \emph{one} time direction, for the sake of simplicity. 

Let us now go back to the details of the model. In order to understand how the "absorbtion" mechanism works, imagine that we have started out from $\tilde{T} (- \infty) = 0$. Then, at a time $t_1$, we have a $\delta$-function "pulse", $T_{\alpha \beta} = k_1 f_{\alpha \beta} (t_1) \delta (t - t_1)$. The evolution equation for $\tilde{T}$ tells us that $\tilde{T}_{\alpha \beta} (t_1^-) = 0$ and $\tilde{T}_{\alpha \beta} (t_1^+) = k_1 f_{\alpha \beta} $. Then, due to the "decay" process, at the time $t_2$ its value will become $\tilde{T}_{\alpha \beta} (t_2) = k_1 f_{\alpha \beta} e^{-k_2 (t_2 - t_1)}$. Since the above picture is clearly linear, if we have a \emph{continuous} source $f_{\alpha \beta}$, the equation becomes
\beq \tilde{T}_{\alpha \beta} (t_0) = k_1 \int f_{\alpha \beta} e^{-k_2 (t_0 - t)} dt \eeq
Now, since we postulated that $T_{\alpha \beta}$ is a source of excitation of $\tilde{T}_{\alpha \beta}$, we can safely replace $f$ with $T$, which gives us
\beq \tilde{T}_{\alpha \beta} (t_0) = k_1 \int T_{\alpha \beta} e^{-k_2 (t_0 - t)} dt \eeq
Thus, if $k_2$ is sufficiently small, this, in fact, becomes an integral over large enough time interval in order for different contributions to $\tilde{T}$ to "average out". Now, as was said earlier, each \emph{individual} contribution has an upper bound, but, due to the fact that they come at different times, they "add up" to something unbounded. As we said before, the "upper bound" of individual portions is enforced by replacing $T$ with $T'$:
\beq \tilde{T}_{\alpha \beta} (t_0) = k_1 \int T'_{\alpha \beta} e^{-k_2 (t_0 - t)} dt \eeq
where, as we said before, 
\beq T'_{\alpha \beta} = M_0 T_{\alpha \beta} {\rm exp} \; \Big(- \sum_{\mu \nu} T_{\mu \nu}^2 \Big) \eeq
As was previously stated, we are forced to violate relativity by having the same sign in exponent, as a result of lightcone singularities that come with any Lorentz-covariant neighborhoods. 

However, we never claimed that our theory is relativistic; we only claimed that it is "local". In fact, we have already violated relativity by having multiple speeds of light, which can be "combined" to produce a single preferred frame. In neither case, the locality of a theory did not suffer. In case of $c_s$ we claimed the locality was preserved because $c_s$ is still finite, even though very large. In case of preferred time direction, we have even stronger case: we don't even envoke "very large" velocities either. Thus, we do not view either of the above as failures of the theory. 

What might be the issue, however, is the possibility that integral of exponent has stochastic jumps. Regardless of how "small" they might be, that would be a problem since it would require us to come up with "another" averaging procedure, which would effectively mean begging the question. At this point I do not have rigourous proof one way or the other, but I hope that such is not the case. 

\subsection*{3.3.3 Modification of Einstein's equation to accommodate small energy non-conservation}

 We know from linear algebra that the existence of soutions and their uniqueness are equivalent: after all, both are linked to the determinant of the matrix being non-zero. In the context of differential equations, "uniqueness" is equivalent to \emph{lack} of gauge freedom. Thus, by adding a term that would break gauge invariance we will restore "uniqueness" and, therefore, "existence", which means that we will have solutions for non-conserved sources, as desired. 

However, while breaking guage invariance might be "necessery", not every way of doing so is "sufficient". After all, it is possible that we break invariance that we are familiar with in favor of some other one, which we won't recognize. Thus, we need some other means of showing whether or not Einstein's equation does, in fact, have solution. One way of doing that is selecting a "preferred" time axis $t$ and moving all of the $\partial^2_t$ terms on the left side of the equation, while leaving $\partial_t$, $\partial_x$, $\partial_t \partial_x$, and so forth, on the right. 

If we think of $t$ as the \emph{only} "dynamical variable" describing the evolution of the metric as well as all of its first and second derivatives \emph{apart from $\partial_t \partial_t$}, then we get a familiar "second Newton's law" problem, and the dependence on $\partial_t$ and $\partial_t \partial_x$ can be simply interpretted as "viscocity". Thus, the existence or absence of the solutions can be easilly shown based on whether or not the determinant of coefficient matrix is zero. 

It is important to point out that $t$-axis is \emph{not} a geodesic. After all, suppose two timelike geodesics, $\gamma_1$ and $\gamma_2$ intersect at two points, $p^{\mu}$ and $q^{\mu}$, where $q^{\mu}$ is "to the future" of $p^{\mu}$. Now, if we know "initial metric and its derivatives" at $p^{\mu}$, we can determine the "final metric and its derivatives" at $q^{\mu}$ by two different methods: we can either track the evolution of metric along $\gamma_1$, or along $\gamma_2$. It is easy to see that if, in both cases, the evolution is completely determined by Einstein's equation, we might as well be "forced" to get different results, unless metric obeys some restrictions to begin with, which, of course, begs the question.

On the other hand, if we think of $t$ as a "flat" coordinate, independent of a metric, then we are guaranteed that the two \emph{lines} along $t$-axis would not intersect, which would remove the ambiguity from our result. While this contradicts the spirit of general relativity, the end result is "covariant". After all any dynamical equation is either invertable or it isn't. The answer to this question should be the same, regardless of the "coordinate frame" we have "chosen" to answer this. 

Since Einstein's equation for arbitrary metric is too complicated to analyze, we will consider a linear approximation. Within this approximation we will first convince ourselves that the equation is not invertible, and then find a way to "modify" it to make it such. The linear approximation to Einstein's equation is 
\beq R_{\mu \nu} - \frac{1}{2} Rg_{\mu \nu} = -\frac{1}{2} \partial^{\alpha} \partial_{\alpha} h_{\mu \nu} + \frac{1}{2} \partial_{\mu} \partial_{\alpha} h^{\alpha}_{\nu} + \frac{1}{2} \partial_{\nu} \partial_{\alpha} h^{\alpha}_{\mu} - \frac{1}{2} \eta_{\mu \nu} \partial_{\alpha} \partial_{\beta} h^{\alpha \beta} - \frac{1}{2} \partial_{\mu} \partial_{\nu} h + \frac{1}{2} \eta_{\mu \nu} \partial^{\alpha} \partial_{\alpha} h \eeq
We will now compute the above expression for all choices of $(\mu \nu)$. However, we will leave out all terms except for the second time derivatives. We will denote the expressions \emph{not} containing second time derivatives by dots. Thus, we get
\beq R_{00} - \frac{1}{2} Rg_{00} = \cdots \; ; \;   R_{11} - \frac{1}{2} Rg_{11} = \frac{1}{2} (\partial_0 \partial_0 h^{22} + \partial_0 \partial_0 h^{33}) + \cdots \nonumber \eeq
\beq R_{10} - \frac{1}{2} Rg_{10} = \cdots \; ; \; R_{12} - \frac{1}{2} Rg_{22} = - \frac{1}{2} \partial_0 \partial_0 h^{12} + \cdots \eeq
The rest of the components are obvious from the permutting indeces. Now, in order to find a dynamics, we have to solve the above equation for $\partial_0 \partial_0 h^{\mu \nu}$. Thus, in order to see whether or not the equation is solvable, we have to compute a determinant of corresponding $10 \times 10$ tensor.  In order to write that tensor in the most transparent form, we will define coordinates of "vectors" it acts upon as follows:
\beq (v_1, \cdots, v_{10}) = \partial_0 \partial_0 (h^{00}, h^{11}, h^{22}, h^{33}, h^{01}, h^{02}, h^{03}, h^{12}, h^{13}, h^{23}) \eeq
In these coordinates, our equation takes the form $A \vec{v} = \vec{w}$, where 
\beq A = \frac{1}{2} \;  {\rm diag} \; (0_{1 \times 1}, B_{3 \times 3}, 0_{3 \times 3}, -I_{3 \times 3}) \eeq
and
\beq B= \left( \begin{array}{ccc} 0 & 1 & 1 \\ 1 & 0 & 1 \\ 1 & 1 & 0 \end{array} \right) \eeq
As a result of $0_{1 \times 1}$ and $0_{3 \times 3}$ components of a diagonal, the determinant is zero. This confirms our earlier prediction that Einstein's equation, as it stands, is not convertible. However, due to the fact that the determinant of $B$ is $2$, it is easy to see that the problem can be cured by replacing $A$ with $A + \epsilon I_{10 \times 10}$. After all, this would replace $0_{1 \times 1}$ and $0_{3 \times 3}$ with $\epsilon I_{1 \times 1}$ and $\epsilon I_{3 \times 3}$, both of which have non-zero determinants. At the same time, by continuity, $B_{3 \times 3} + \epsilon I_{3 \times 3}$ will have determinant close to $2$ and, therefore, also non-zero. 

Now, adding this extra term is equivalent to adding $\epsilon \partial_0 \partial_0 h^{\mu \nu}$ to our equations. Since we would like to have a covariant theory, we instead add a term $\epsilon \partial^{\partial^{\alpha}} \partial_{\alpha} h_{\mu \nu}$. Furthermore, we notice that derivatives of $\eta_{\mu \nu}$ are $0$ and,  therefore, we can freely replace $\epsilon \partial^{\alpha} \partial_{\alpha} h_{\mu \nu}$ with $\epsilon \partial^{\alpha} \partial_{\alpha} (\eta_{\mu \nu} + h_{\mu \nu})$. Finally, in order to generalize our equation to arbitrary metric, we replace $\eta_{\mu \nu} + h_{\mu \nu}$ with $g_{\mu \nu}$, which gives us
\beq R_{\mu \nu} - \frac{1}{2} Rg_{\mu \nu} + \epsilon \partial^{\alpha} \partial_{\alpha} g_{\mu \nu} = T_{\mu \nu} \eeq
It is important to notice that, in light of $\partial_{\alpha}$ being used in place of $\nabla_{\alpha}$, the $\epsilon$-term in the above equation is \emph{not} covariant. In fact, in light of the fact that $\nabla_{\alpha} g_{\mu \nu} =0$, we couldn't have possibly made it covariant even if we wanted to. But, like was mentioned earlier, violation of covariance is one of the necessary steps of turning non-invertible equation into the invertible one. This expression corresponds to a Lagrangian
\beq S = \int d^4 x \; \sqrt{-g} \; \Big( R+ kT + \frac{\epsilon}{2} \; \partial^{\alpha} g^{\mu \nu} \partial_{\alpha} g_{\mu \nu} \Big) \eeq
where $\epsilon$-term likewise breaks the general relativistic covariance. 

It is important to admit, however, that in the above analysis we were working exclusively in the linear approximation. For the general metric, it is quite possibe for determinant to "accidentally" become zero. In fact, if there are two timelike-separated points, $p^{\mu}$ and $q^{\mu}$, and the determinant has opposite sign at these two points, then \emph{any} curve $\gamma$ that connects them will have some point $r_{\gamma}$ at which the determinant is $0$. If we now alter the curve $\gamma$, then the collection of such points will form a surface, and the determinant will be zero \emph{everywhere} on that surface!

The good part is that the surface is "infinitely smaller" than anything three-dimensional. The bad news, however, is that in a small-but-finite vicinity of a surface, the second time derivatives of gravitational field will undergo unwanted large-but-finite variation. This will have large unwanted impact throughout the three-dimensional region "to the future" of that surface. Thus, in order to avoid this, we have to "tame" the behavior of time derivatives somehow.  

As a "quick fix", we are going to single out "preferred" $t$-direction since, for the purposes of avoiding singularity, it \emph{only} necessary to tame $\partial_t \partial_t$-terms. When we first introduced "preferred frame" few paragraphs earlier, we have stated that the choice of that frame has no impact on our conclusion of whether or not a differential equation is convertible. However, now that we are trying to do \emph{modification} of an equation in our "preferred frame", this \emph{does} lead to a prediction of different results depending on that frame; although, of course, the difference will be of the order of magnitude of "small correction" we attempt to introduce and, therefore, undetectable. 

Of course, for the future, it is important to explore more covariant ways of modifying gravity. But, just for the purposes of this paper, we will settle on violating relativity; our excuse being that we have already done that anyway when we have introduced Pilot Wave model. As we said before, we can write the second time derivatives in a form of a single vector $\vec{v} \in \mathbb{R}^{10}$, which allows us to write the dynamics as $\vec{v} = A^{-1} \vec{w}$, for the tensor $A$ specified earlier. We will now replace it with 
\beq \vec{v} = \Big(\frac{\tan^{-1} \; (\xi \; det \; A^{-1})}{\xi \; det \; A^{-1}} \Big)^{1/10} A^{-1} \vec{w} \eeq
where $\xi$ is a very small number. In light of the smallness of $\delta$, the overall coefficient is very close to $1$ \emph{as long as} the determinant of $A^{-1}$ does not blow up. However, once it does, the coefficient approaches zero. It is easy to see that the determinant of the "modified" version of $A^{-1}$ approaches $\pi / 2 \delta$, which is small but finite. Thus, if the thickness of a "problematic" region is much smaller then $\xi$, it does not have any detectible effects. 

In fact, we can do "even better" by removing $1/10$ from the power of the above coefficient. In this case, near the singular region the second time derivatives will approach zero, which is just \emph{the opposite} of the issue we were trying to avoid! We have to be careful if we want to do that, because this might potentially "freeze" the singularity once it is reached. What comes to our rescue is that the \emph{first} time derivatives continue to be non-zero. The space variation of the latter will eventually allow us to "get out" of the singular region. But, of course, this needs to be investigated more closely. 

\subsection*{4. Conclusion}

In this paper we have proposed a model in which quantum phenomena are reproduced by means of superluminal, but finite, speed of signals. The difference between"superluminal" and "trully infinite" is a key idea of the model. While we embrace the former, we reject the latter. Thus, we get all the "benefits" of quantum non-locality, without having to pay the price of a very counter-intuitive notions.

The key idea of this paper is that there is no such thing as "configuration space". Instead, the same-charge particles are so strongly correlated that, up to a very good approximation, they \emph{act} like a single point in a configuration space. That correlation is due to them communicating signals with each other, with emission and absorbtion frequencies corresponding to their charges. 

The letter $\psi$, which usually denotes probability amplitude, is now viewed as the internal degree of excitation of particles. But, due to strong correlation, the dynamics forces $\psi$ to be nearly the same among particles with the same "charge". As a result, we wrongly interpret $\psi$ as a parameter on configuration space or Fock space. 

According to proposed model, the particles do not get greated or annihilated. When we make a "switch" from one state to another, what happens is that some partiles stop being highlighted while others begin to be. They are "highlighted" through another internal degree of freedom, $B$, which is also subject to similar global correlations. The values of $B$ define observed reality, as particles with nearly-zero $B$ are "invisible". 

Finally, gravity is introduced as a "classical" field that is coupled to the \emph{end product} of the above process (namely $B$). Until $B$ is produced, the only role $g_{\mu \nu}$ plays is defining a \emph{fixed} geometrical background for the non-gravitational processes. Then, after $B$ is produced, $g$ "looks" at it and "decides" how it wants to alter; its aterations are, of course, identified with the "gravitatioal effects" of $B$. 

However, the energy-momentum "source" produced by $B$ happened not to be conserved. Thus, there is no exact solution of Einstein's equation with that source. Since we would like to view $g_{\mu \nu}$ as classical, we would like it to be exact solution of \emph{some} equation. Thus, we have modified Einstein's equation by adding to it small kinetic term. While we have not proven the existence of solution in a general case, we have at least made a good argument in favor of solution being available in most "regular" cases. We then attempted some maneurs of getting around the problems in more exotic situations. 

This model should be compared to an earlier attempt of comming up with superluminal locality, \cite{completely} (since I happened to be the author of that attempt as well, I would like to take a liberty and criticize my own work). In that other attempt, instead of introducing correlations between particles, I have introduced correlations between \emph{regions} of space into which particles \emph{happened} to fall. I believe that was a lot less elegant than the current work. After all, back then I had to introduce oscillations of points in space \emph{not} occupied by particles. This amouts to a lot of extra unnecessary information, which fails Ocam's razor. 

Apart from this, the "structure" of the "regions" introduced in \cite{completely} was very complicated. Thus, it would be difficult to combine that model with gravity since the latter is likely to "disturb" a very complicated structure that was employed. While the current approach also has some problems when it comes to combining it with gravity, these problems are all unavoidable, as compared to the extra complications introduced in the previous approach that could have easilly been avoided.

The gravitational part of this work should be compared with \cite{cheat} (which was also authored by me). In that paper we have also attempted to describe the production of gravity by non-conserved sources, for the same reasons as we did it here. However, again, the approach of that work was a lot less natural than what we are doing now. In \cite{cheat} we came up with a model where gravitational field performs "trial and error" method as it attempts different "changes" that would bring it closer to being a solution of Einstein's equation. Even though I successfully described trial-and-error through a set of differential equations, it still feels a bit artificial. On the other hand, in the current paper it was done in a lot more natural way, through

\end{document}